\documentclass[numberedappendix,appendixfloats]{emulateapj}
\usepackage{amsmath,graphicx}

\shortauthors{K. Omukai et al.}
\shorttitle{Low-Metallicity Star Formation}
 
\begin{document}
 
\title{Low-Metallicity Star Formation:\\ Prestellar Collapse and
Protostellar Accretion in the Spherical Symmetry} 

\author{Kazuyuki Omukai \altaffilmark{1, 2}, 
Takashi Hosokawa \altaffilmark{1, 2, 3}
and Naoki Yoshida \altaffilmark{4}} 
\altaffiltext{1}{Department of
Physics, Kyoto University, Kyoto 606-8502, Japan}
\altaffiltext{2}{Division of Theoretical Astronomy, National
Astronomical Observatory, Mitaka, Tokyo 181-8588, Japan}
\altaffiltext{3}{Jet Propulsion Laboratory, California Institute of
Technology, Pasadena CA 91109, USA} \altaffiltext{4}{IPMU, University
of Tokyo, Kashiwa, Chiba 277-8583, Japan}
\email{omukai@tap.scphys.kyoto-u.ac.jp}

\begin{abstract}
The collapse of dense cores with metallicities $0-1Z_{\sun}$ is
studied by hydrodynamical calculations coupled with detailed chemical
and radiative processes.  For this purpose, we construct a simple
chemical network with non-equilibrium reactions among 15 chemical
species, ${\rm H^+}$, ${\rm e}$, ${\rm H}$, ${\rm H_2}$, ${\rm D^+}$,
${\rm D}$, ${\rm HD}$, ${\rm C^+}$, ${\rm C}$, ${\rm CO}$, ${\rm
CO_2}$, ${\rm O}$, ${\rm OH}$, ${\rm H_2O}$, and ${\rm O_2}$, which
reproduces the abundance of important molecular coolants by more
detailed network very well.  Starting from the initial density of
$10^{4}{\rm cm^{-3}}$, the evolution is followed until the formation
of a hydrostatic protostar at the center $\sim 10^{21}{\rm cm^{-3}}$.
In a lower-metallicity gas cloud, the temperature during the collapse
remains high owing to less efficient cooling. After the cloud core
becomes optically thick to the thermal emission by dust, temperature
evolution at the center converges to a single trajectory, except for
cases with metallicity $\leq 10^{-6}Z_{\sun}$, where the temperature
remains slightly higher than in higher metallicity cases even after
becoming optically thick to thermal radiation by the H$_2$
collision-induced emission. The protostellar masses at their formation
are a few $10^{-3}M_{\sun}$, being slightly higher for cases with
$\leq 10^{-6}Z_{\sun}$.  Using the temperature evolution at the center
as a function the density, we discuss the possibility of fragmentation
during the dust-cooling phase. The critical metallicity for the
fragmentation is $10^{-5}Z_{\sun}$ assuming moderate elongation of the
cloud cores at the onset of this phase. From the density and velocity
distributions at the time of protostar formation, we evaluate the mass
accretion rate in the subsequent accretion phase.  The accretion rate
is larger than the Shu accretion rate for the inside-out collapse from
an initially static cloud $\simeq c_{\rm s}^3/G$, where $c_{\rm s}$ is
the sound speed in the prestellar gas, by about a factor of ten owing
to more dynamical nature of the collapse.  Using these accretion
rates, we also calculate the evolution of the protostars under the
assumption of stationary accretion flow.  For $\geq 10^{-4}Z_{\sun}$,
we succeed in following their evolution well after the arrival to the
main-sequence phase. For lower metallicity cases, however, owing to
too high accretion rates $\ga$ a few $10^{-3} M_{\sun}$/yr, the total
luminosity, which consists of contribution from accretion and internal
one, reaches the Eddington limit, thereby rendering the stationary
accretion impossible for $\ga 100M_{\sun}$. Finally, we discuss
possible suppression of fragmentation by heating of the ambient gas by
protostellar radiation, which is considered important in the
contemporary star formation.  We argue that it is negligible for
$<10^{-2}Z_{\sun}$.

\end{abstract}
 
 
\section{Introduction}
The first stars in the universe are thought to be formed from a
chemically pristine gas. Recent theoretical studies and cosmological
simulations suggest that these stars are very massive, perhaps even
greater than 100 solar masses (Bromm, Coppi \& Larson 1999; Abel,
Bryan \& Norman 2002; Yoshida, Omukai \& Hernquist 2008; Ohkubo et al. 
2009).  Stars in the solar neighborhood are, on the other hand,
typically of low-mass with the initial mass function peaking around
$\sim 0.3M_{\sun}$ (e.g., Kroupa 2002; Bastian et al. 2010).  Although
the characteristic mass of old halo stars -- Population II stars-- is
not well known, a substantial fraction of them were formed with low
masses and thus are still observed in the Galaxy. Clearly, there must
have been a transition in the characteristic stellar mass from massive
to low-mass ones.

Stars are formed under rather different conditions in the early
universe and in the local universe.  For example, metallicity of the
interstellar gas, strength of a radiation field, existence of cosmic
rays and a magnetic field, are all thought to be critical quantities
that affect the process of star formation.  Gas metallicity appears to
be among the most important quantities that affects the thermal
evolution of a prestellar gas and hence might determine the
characteristic stellar (at least gas cloud) mass (e.g., Bromm et
al. 1999; Omukai 2000; Schneider et al. 2002).

Observationally, no clear differences are found in the stellar initial
mass function (IMF) for regions with different metallicities in the
local universe (e.g., Kroupa 2002; Bastien, Covey, \& Meyer 2010).
However, the difference of metallicity in the examined regions is only
a factor of a few, and thus it remains unclear if the characteristic
stellar mass can change drastically with much lower metallicity.
Interestingly, there is an observational suggestion that very
metal-poor stars in the halo had larger characteristic mass (Lucatello
et al. 2005).  The carbon-enhanced metal-poor stars are considered to
be secondary, i.e., lower-mass, stars in binaries. The other star in
the binary is typically an intermediate-mass star, whose mass is a few
solar masses.  The fraction of carbon-enhanced stars increases with
decreasing metallicity, indicating a higher fraction of
intermediate-mass stars than in the solar neighborhood.

Observations show that, in the Galaxy, the shape of dense-core
mass-function resembles that of the stellar IMF (e.g., Motte et
al. 1998).  This fact indicates that the characteristic stellar mass
is originally imprinted through the formation process of the dense
cores, e. g., fragmentation of star-forming gas.  According to recent
detailed numerical simulations, the frequency of fragmentation depends
strongly on the temperature evolution of gas. Fragmentation occurs
frequently as long as the effective ratio of specific heat $\gamma =
{\rm dlog} p/ {\rm dlog} \rho $, where $p$ is the pressure and $\rho$
is the density of the gas, is less than unity, i.e., the temperature
decreases with increasing density during the collapse. With $\gamma
>1$, on the other hand, the fragmentation hardly occurs (Li et
al. 2003). The moment when $\gamma$ exceeds unity, i. e., at a
temperature minimum, marks the end of the fragmentation epoch.  The
Jeans mass at this instant is imprinted as a characteristic
fragmentation scale.  Those numerical results can be interpreted by
dynamical fragmentation theory of non-isothermal gas (Larson 1985),
which predicts that the core mass-scale is set by fragmentation of
filamentary clouds into round cores, which occurs when the
isothermality breaks.  In the first star formation, dense cores are
indeed produced around the temperature minimum at $\sim 10^{4}{\rm
cm^{-3}}$, where the H$_2$-cooling rate saturates owing to the
collisional de-excitation.  Also, for the present-day star formation,
the Jeans mass at the dust-gas coupling, where the temperature becomes
minimum, falls on the correct range for the characteristic stellar
mass (Larson 1985; 2005, Whitworth et al. 1998).

Prestellar evolution of both present-day and primordial clouds has
already been modelled in detail.  In a primordial gas cloud, important
processes are the chemistry and radiative processes of H$_2$, which
are simple enough to incorporate into numerical simulations in a
direct, self-consistent manner (Omukai \& Nishi 1998; Abel, Bryan \&
Norman 2002; Yoshida, Omukai \& Hernquist 2008). The situation is
clearly different for the present-day star formation. Quite a large
number of radiative and chemical processes are operating in a
metal-enriched gas cloud, e.g., cooling by atomic and molecular line
emission, heating by the dissipation of turbulence and by external
radiation fields, in particular, in a diffuse medium.  At first
glance, this complexity makes it extremely hard to model the
presente-day star formation process.  In addition, once a central
protostar turned on, the heating of ambient gas by its radiation
becomes important (Krumholz 2006; Bate 2009).  However, considering
the fact that the cooling by dust grains dominates other radiative
cooling processes after the formation of dense cores ($\ga 10^{4}{\rm
cm^{-3}}$) and before the protostar formation, a simplification is
possible: the prestellar collapse of dense cores can be modelled by
considering only the radiative transfer of dust thermal emission.
This makes the chemistry and radiative process effectively decouple
(Gaustad 1963; Hattori, Nakano, \& Hayashi 1969), and thus detailed
calculations have been done since long ago (Larson 1969; Narita,
Nakano, \& Hayashi 1970; Winkler \& Newman 1980; Masunaga \& Inutsuka
2000).

In Pop II star formation, the main difficulty lies in the coupling of
chemical reactions and radiative cooling processes: one needs to solve
chemical reactions to know the abundances of coolant species
accurately. Miki \& Nakano (1975) and Yoshii \& Sabano (1980) studied
the thermal evolution of collapsing low-metallicity clouds using
one-zone models.  Similarly, Bromm \& Loeb (2003) and Santoro \& Shull
(2006) examined the role of fine-structure line cooling in a
low-metallicity cloud.  These studies, however, do not follow chemical
reactions but assume that metals in the gas phase remain in the atomic
phase.  Omukai (2000) and Omukai et al. (2005; hereafter O05) solved
the detailed chemical reactions as well as radiative processes. They
identified the importance of cooling by CO, OH and H$_2$O for
metallicity $Z < 10^{-1}Z_{\sun}$.  There still remains overall
uncertainties, however.  The dynamics is affected by the thermal
evolution via pressure force, so the treatment in the one-zone models
has not yet been justified. Three-dimensional hydrodynamical
calculations were also performed by several authors (Bromm et
al. 1999; Smith \& Sigurdsson 2007; Smith et al. 2009). A slightly
different approach is to use a prescribed effective equation-of-state
in hydrodynamical calculations.  In such studies, temperature
evolution is given solely as a function of density, based on the
result of one-zone models (Clark, Glover \& Klessen 2008; Tsuribe \&
Omukai 2006, 2008, Machida et al. 2008; 2009).  Jappsen et al. (2007;
2009a,b) included chemical reactions self-consistently in their
hydrodynamical calculations, but followed only early evolution phase
with density $\la 10^{4}{\rm cm^{-3}}$. None of these calculations
provide a complete picture of prestellar gas collapse with
hydrodynamics, chemistry, and radiative transfer.

In this paper, we study the collapse of star-forming dense cores with
different metallicities using radiation hydrodynamics calculations
including chemical reactions for important molecular coolants.  We
follow their evolution up to the formation of protostars at the center
under the assumption of spherical symmetry.  The structure of the
cores at the moment of the protostar formation sets the initial
conditions for the subsequent accretion phase of the protostar. In
particular, it determines the protostellar accretion rate.  Using the
obtained accretion rates, we then calculate the evolution of
protostars until they start hydrogen burning and arrive at the main
sequence phase.

This paper is organized as follows. In Sec. 2, the numerical method is
described. We present the prestellar evolution of dense cores in
Sec. 3 and then the protostellar evolution in the accretion phase in
Sec. 4.  Sec. 5 is for discussion. Finally, we summarize the paper in
Sec. 6.  In Appendix A the chemical network solved is presented.
Appendix B contains the fitting parameters for the OH and CO cooling
functions.

\section{Method of Calculation}
We calculate the gravitational collapse of the dense cloud cores with
metallicities $Z=0-1~Z_{\sun}$ by solving the radiation transfer and
chemical reactions as well as hydrodynamics under spherical symmetry.
Starting from the initial density $\sim 10^{4}~{\rm cm^{-3}}$ at the
center, we follow the evolution until shortly after the formation of a
protostar at $\sim 10^{21} {\rm cm^{-3}}$.

\subsection{Hydrodynamics}

We solve hydrodynamics using the Lagrangian difference method with the
standard von Neumann-Richtmyer viscosity (see, e.g., Mihalas \&
Weibel-Mihalas 1984).  The results presented here are obtained with
200 grid points logarithmically spaced at the initial moment, with the
innermost cell containing a mass of $10^{-6}M_{\sun}$.  The outer
boundary is fixed at its initial position.  To see the numerical
convergence of our results, we also carry out calculations for the
same setting but with the number of the cells halved.  We have
confirmed that the results agree well.

\subsection{Cooling Processes}

We include the same cooling processes as in O05 with some updates of
the coefficients.  The total cooling rate $\Lambda$ consists of three
processes, i.e., cooling via atomic and molecular line photons,
cooling by continuum radiation, and the chemical cooling / heating:
\begin{equation}
\Lambda = \Lambda_{\rm line} + \Lambda_{\rm cont} + \Lambda_{\rm
chem}.
\end{equation} 
When the chemical heating exceeds the cooling, the last term
$\Lambda_{\rm chem}$ becomes negative.

\subsubsection{Line-emission Cooling}
The line-cooling processes included are those by fine-structure
transitions of CII, CI, and OI, and rotational transitions of H$_2$,
HD, CO, OH, and H$_2$O. For H$_2$, vibrational transitions are also
considered.  Atomic and molecular line photons are trapped in the core
when the optical depth is large. This effect is taken into account by
reducing the cooling rate by a emission line multiplying the escape
probability $\beta_{\rm L}$;
\begin{equation}
\beta_{\rm L}=\frac{1-e^{-\tau_{\rm L}}}{\tau_{\rm L}} e^{-\tau_{\rm
C}},
\end{equation}
where $\tau_{\rm L}$ is the optical depth of the line and $\tau_{\rm
C}$ is that of continuum.  The optical depth is evaluated with the
radially-integrated column density
\begin{equation}
N_{\rm X}(r)= \int_{r} y(X) n_{\rm H} dr',
\end{equation}
where $y(X)$ is the concentration of the emitting species $X$, and
$n_{\rm H}$ is the number density of hydrogen nuclei. The above
integral is over the radial interval where the velocity shift is
within the line width; $|v(r')-v(r)|<(v_{\rm D}(r')+v_{\rm
D}(r))/2$. In Omukai \& Nishi (1998), transfer of H$_2$ lines is
solved not only by using correct averaging over solid angles but also
by resolving frequency distributions within lines.  By comparing the
results with the same initial condition as Omukai \& Nishi (1998) but
with our approximations for line radiation transfer, we have confirmed
that the results agree very well.

The coefficient for CO, OH and H$_2$O line transitions are updated
from O05. For H$_2$O, we use the cooling rate given by Neufeld \&
Kaufman (1993) for $T > 100$K, and Neufeld, Lepp, \& Melnick (1995)
for $T < 100$K. For CO and OH, we calculated the cooling rates as in
O05 by using the transition coefficients from Leiden Atomic and
Molecular Database (Schoier et al. 2005) and fit the results by the
same form of the formula as in Neufeld \& Kaufman (1993).  The fitting
parameters for the CO and OH cooling rates are summarized in Appendix
A.

\subsubsection{Continuum-emission Cooling}
We solve transfer of continuum radiation by the variable Eddington
factor method with grey approximation (Mihalas \& Weibel-Mihalas
1984).  For a given mean intensity $J$, the dust temperature $T_{\rm
d}$ is determined by the energy balance on the dust grains:
\begin{equation}
\kappa_{\rm d}B(T_{\rm d})=\kappa_{\rm d}J+\Gamma_{\rm coll},
\end{equation}
where the Planck function $B(T)=\sigma_{\rm B}T^{4}/\pi$, $\kappa_{\rm
d}$ is the Planck mean opacity of dust and the heating rate for the
dust by collisions with gas particles $\Gamma_{\rm coll}$ is given by
(Hollenbach \& McKee 1979)
\begin{equation}
\Gamma_{\rm coll}=4.4 \times 10^{-6} (f/\rho)_{\rm gr} n_{\rm H}
\left( \frac{T}{1000{\rm K}} \right)^{1/2} (T-T_{\rm d}),
\label{eq:Gcoll}
\end{equation}
where $T$ is the gas temperature, and $(f/\rho)_{\rm gr}$ is total
volume of the dust per unit gas mass, which is taken from Pollack et
al. (1994).  For given gas and dust temperatures, the source function
can be written as
\begin{equation}
S=\frac{\kappa_{\rm d}B(T_{\rm d})+\kappa_{\rm g}B(T)} {\kappa_{\rm
d}+\kappa_{\rm g}},
\end{equation}
where $\kappa_{\rm g}$ is the gas continuum opacity.  Once
distribution of the source function $S$ is given, the mean intensity
$J$ is found by carrying out standard ray-tracing method.  We do
iterations until the consistent distributions of $T_{\rm d}$ and $J$
are obtained.  Once radiative transfer is solved, the cooling rates by
dust and gas continuum emissions are given by
\begin{equation}
\Lambda_{\rm d} = 4\pi \kappa_{\rm d} [B(T_{\rm d})-J]
\end{equation}
and
\begin{equation}
\Lambda_{\rm g} = 4\pi \kappa_{\rm g} [B(T)-J],
\end{equation}
respectively.

\subsubsection{Chemical Cooling and Heating}
The change in chemical binding energy works as heating or cooling of
the gas. The heating and cooling associated the chemical reactions are
treated in the same way as in Omukai (2000). We note here that
important processes are heating/cooling by the H$_2$
formation/dissociation, respectively.

\subsection{Chemical Processes}
We solve the reduced chemical network of O05, which correctly
reproduces the evolution of major coolant species obtained by a more
detailed network calculation.  Our network includes 24 compounds of
hydrogen, deuterium, carbon and oxygen.  For the following 15 major
species, non-equilibrium reactions are solved: ${\rm H^+}$, ${\rm e}$,
${\rm H}$, ${\rm H_2}$, ${\rm D^+}$, ${\rm D}$, ${\rm HD}$, ${\rm
C^+}$, ${\rm C}$, ${\rm CO}$, ${\rm CO_2}$, ${\rm O}$, ${\rm OH}$,
${\rm H_2O}$, ${\rm O_2}$ (see Table 1).  The other 9 minor species,
i.e., ${\rm H^-}$, ${\rm CH}$, ${\rm CH_2}$, ${\rm CO^+}$, ${\rm
O^+}$, ${\rm OH^+}$, ${\rm H_2O^+}$, ${\rm H_3O^+}$, ${\rm O_2^+}$,
which appear only as intermediaries of the reactions among the above
major species, are set to be chemical equilibrium (see Appendix B for
more details).  By performing some experiments using the one-zone
models, we confirmed that the evolution of the major species'
abundances in this method agree very well with that in the full
non-equilibrium treatment.

\subsection{Initial Core Models}

\begin{figure}
\plotone{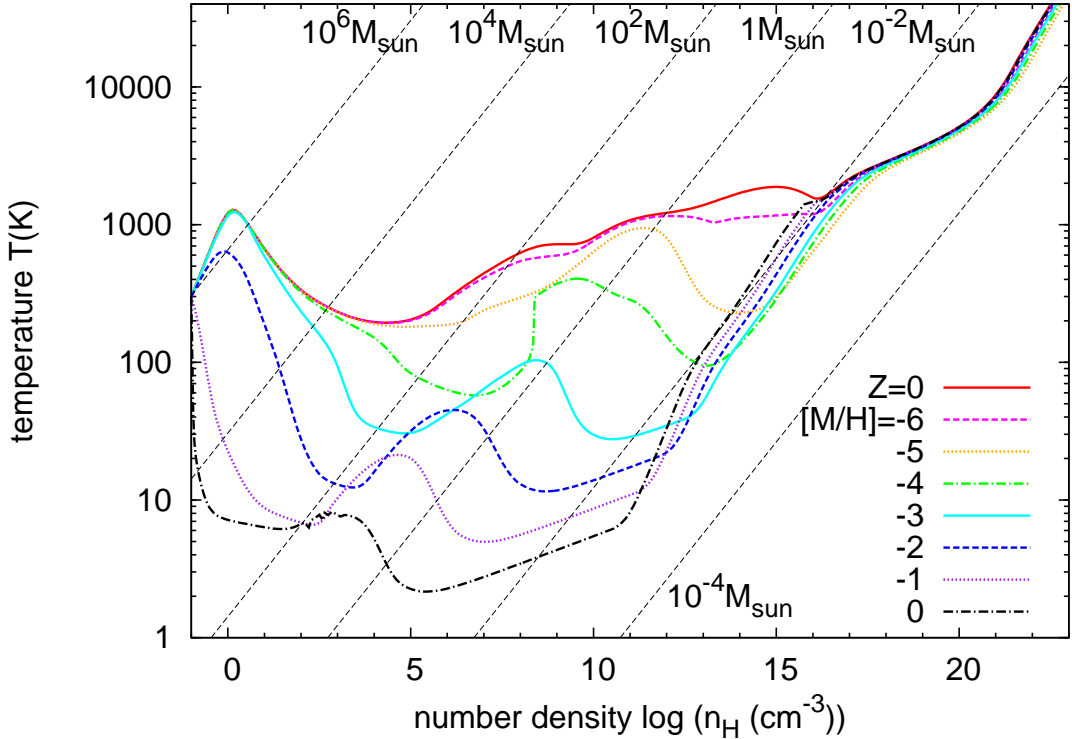} 
\caption{The evolution of temperatures
in prestellar cloud cores with metallicities $Z/Z_{\sun}=0, 10^{-6},
10^{-5}, 10^{-4}, 10^{-3}, 10^{-2}, 10^{-1},$ and 1, as functions of
the number density, which is calculated by one-zone models. The dashed
lines indicate the constant Jeans masses.  For those above
$10^{2}M_{\sun}$ (below $1M_{\sun}$), the gas is assumed to be fully
atomic (molecular) in drawing those lines.
\label{fig:one-zone}}
\end{figure}

Initial models for the hydrodynamical calculations are prepared using
the early evolution in our one-zone model.  The model is the same as
that developed in O05, but some rate coefficients of radiative and
chemical processes are updated to be the same as in the hydrodynamical
models.  Figure \ref{fig:one-zone} shows the prestellar temperature
evolution calculated by the one-zone model with different
metallicities.  The temperature and chemical abundances of the initial
models are assumed to be radially constant, and their values are taken
from the one-zone models at $10^{4}{\rm cm}^{-3}$. For density and
velocity distributions, we assume the critical Bonner-Ebert sphere at
rest, which is the hydrostatic configuration on the verge of the
collapse.  To initiate gravitational collapse, we enhanced the density
by twenty percent.  The central number density is $1.2 \times 10^{4}
{\rm cm^{-3}}$ at the beginning of the hydrodynamical calculations.
In addition to the case of primordial gas, $Z=0$, we study the cases
every one dex in metallicity from $Z=10^{-6}Z_{\sun}$ ([M/H]=-6) to
$1Z_{\sun}$ ([M/H]=0).  The total masses of the dense cores depend on
the initial temperature at $1.2 \times 10^{4} {\rm cm^{-3}}$ for the
hydrodynamical calculations; 4.7 ([M/H]=0), 37 (-1), 52 (-2),
$1.7\times 10^{2}$ (-3), $1.6\times 10^{3}$ (-4), $2.3\times 10^{3}$
(-5), and $2.4\times 10^{3}M_{\sun}$ (-6, and $Z=0$).  The outer
boundary is fixed at its initial position.

\section{Prestellar Collapse}

Figure \ref{fig:nTc.all} presents the temperature evolution at the
center of the prestellar cores as a function of the number density.
The overall evolution is quite similar to that calculated by the
one-zone model (Fig.\ref{fig:one-zone}), justifying the one-zone
treatment for the core evolution.  There are, however, small
disagreements, in particular, at high densities and for low
metallicity cases.  We defer detailed discussion on these differences
to later in Sec. 3.4, but here describe which thermal processes
control the temperature evolution at each metallicity.  The
contribution to the cooling and heating rates by individual processes
are presented in Figure \ref{fig:cool.all} for different
metallicities.  This should be compared with Figure 2 of O05, where
similar plots for the one-zone models are presented.  In Figure
\ref{fig:gamma}, the effective ratio of specific heat at the center
$\gamma=d{\rm ln} p/ d{\rm ln} \rho$, which tells the variation of
pressure in response to the density variation, is shown for those
cases.  Note that $\gamma-1$ equals the gradient of the curve in
Figure \ref{fig:nTc.all} for constant molecular weight. The effective
ratio of specific heat is an important index to examine dynamical
response of self-gravitating clouds to thermal evolution.  For
example, the clouds easily fragments as long as $\gamma <1$, while
fragmentation is strongly prohibited for $\gamma >1$ (Li et al. 2003).
Another critical value is $\gamma=4/3$. If $\gamma$ exceeds this
value, the dynamical collapse is halted as the pressure overcoming the
gravity, and a hydrostatic object is formed.

\subsection{Thermal Evolution in the Metal-free Case}
In this section, we review thermal evolution of the cloud core of a
metal-free gas.  We then describe the effects of metallicity later in
Sec. 3.2.  We focus on deviations from the metal-free case.  In the
case of metallicity [M/H]=-6, metallicity effects are so small that
the temperature evolution is almost identical to the metal-free one
except a slight offset at highest densities ($\ga 10^{20}{\rm
cm^{-3}}$).

Let us summarize here the formation processes of H$_2$, which play a
crucial role in the thermal evolution.  The evolution of H$_2$
concentrations is presented in Figure \ref{fig:nH2}, along with those
in the cases with metals.  Below $\sim 10^{8}{\rm cm^{-3}}$, H$_2$ is
formed by the H$^{-}$ channel:
\begin{eqnarray}
{\rm H + e} &\rightarrow& {\rm H^{-} + \gamma} \\ {\rm H^{-} + H}
&\rightarrow& {\rm H_2 + e},
\end{eqnarray}
catalyzed by a small amount of remaining electrons.  With their
recombination proceeding, the H$^{-}$ channel is quenched and the
amount of formed H$_2$ saturates at $\sim 10^{-3}$
(Fig. \ref{fig:nH2}).  After this plateau, the H$_2$ abundance begins
to increase again at $\sim 10^{8}{\rm cm^{-3}}$ via the three-body
H$_2$ formation:
\begin{equation}
{\rm 2 H + H} \rightarrow {\rm H_2 + H},
\label{eq:3body}
\end{equation}
and
\begin{equation}
{\rm 2 H + H_2} \rightarrow {\rm H_2 + H_2}.
\end{equation}
All the hydrogen is converted to the molecular form via this channel
by the density $\sim 10^{11}{\rm cm^{-3}}$.

Next, let us see the cooling and heating processes
(Fig. \ref{fig:cool.all} a).  Until very high density $\sim
10^{19-20}{\rm cm^{-3}}$ is reached, cooling and heating are always
almost balanced, so that the evolution is nearly isothermal with
temperature differing only by a small factor whereas density increases
by many orders of magnitudes.  The effective ratio of specific heat
$\gamma$ remains below 4/3 but is above 1 in this period except for
brief intervals around $10^{9}{\rm cm^{-3}}$ and $10^{11}{\rm
cm^{-3}}$, where $\gamma$ falls slightly below unity
(Fig. \ref{fig:gamma} a).  The heating is owing to the compression,
but for $10^{9}-10^{12}{\rm cm^{-3}}$, where the H$_2$ formation
heating associated with the three-body reaction (eq. \ref{eq:3body}
below) dominates. For the cooling, the H$_2$-line emission contributes
most until $\sim 10^{13}{\rm cm^{-3}}$, although some lines become
optically thick at $\sim 10^{11}{\rm cm^{-3}}$ and this suppresses the
cooling rate gradually towards a higher density.  The steep decline of
the H$_2$ line-cooling rate at $10^{16}{\rm cm^{-3}}$ is due to the
H$_2$ collision-induced continuum absorption.  Another molecular
species in the metal-free gas, HD, is known to play an important role
in cooling if a metal-free gas is once ionized (Uehara \& Inutsuka
2000; Nagakura \& Omukai 2006; Greif \& Bromm 2006; Yoshida, Omukai,
\& Hernquist 2007; McGreer \& Bryan 2008). In our case, however, it
only contributes comparably to H$_2$ at a brief period at $\sim
10^{4}{\rm cm^{-3}}$.

With gradual increase of temperature, the balance of chemical
equilibrium between the H$_2$ formation (eq. \ref{eq:3body}) and its
inverse dissociation reactions shifts in favor of dissociation in the
range $10^{13}-10^{15}{\rm cm^{-3}}$.  The cooling by this subtle
dissociation is enough to compensate the compressional heating,
because of the large binding energy of molecular hydrogen, $\chi =
4.48$ eV.  At $\sim 10^{15}{\rm cm^{-3}}$, H$_2$ formation is somewhat
resumed owing to the cooling by the H$_2$ collision-induced continuum
emission, which is a molecular analogue of free-free emission.  At
densities greater than $\sim 10^{16}{\rm cm^{-3}}$, the cooling rate
by this process declines rapidly when the central part of the core
becomes optically thick to this same continuum.  This time, with
increase of temperature, the H$_2$ begins to dissociate. The
dissociation cooling balances with the compressional heating until
$\sim 10^{20}{\rm cm^{-3}}$, where the dissociation is almost
completed.  For higher densities, without efficient cooling process,
the temperature increases almost adiabatically and $\gamma$ exceeds
4/3. This marks the end of the prestellar collapse and the formation
of a hydrostatic object, i.e. a protostar, at the center (see
Sec. 3.3).

\subsection{Metallicity Effects on Thermal Evolution}
Addition of metals alters the thermal evolution of star-forming cloud
cores significantly in the following ways.

\subsubsection{Atomic and Molecular Metal-line Emission}

For a metal-free cloud core, the efficient radiative cooling is solely
by H$_2$, i.e. its rovibrational line emission and collision-induced
emission, with some contribution from HD around $\sim 10^{4}{\rm
cm^{-3}}$.  With metal enrichment, the cooling by line emission of
metallic atoms ([CII], [CI], [OI]) and molecules (CO, OH, H$_2$O)
affects the thermal evolution.

Oxygen-bearing molecules, OH and H$_2$O, act as efficient coolants.
In the case of the metallicity [M/H]=-5, since the gas temperature
remains high at $> 200$K, OH molecules formed at $\sim 10^{7}{\rm
cm^{-3}}$ are readily converted to H$_2$O.  Cooling by H$_2$O
rotational transition exceeds H$_2$ cooling in the range
$10^{8}-10^{10} {\rm cm^{-3}}$. With more metals of [M/H]=-(3-4), the
lower temperature conditions delay the conversion of OH to H$_2$O
molecules.  The line emission by OH molecules dominates cooling at
$10^{6}-10^{8}{\rm cm^{-3}}$ before the H$_2$-formation heating raises
the temperature and promotes the conversion to H$_2$O molecules.
After that, H$_2$O significantly contributes to cooling until
$10^{10}{\rm cm^{-3}}$ ($10^{9}{\rm cm^{-3}}$) at [M/H]=-4 ([M/H]=-3,
respectively), when the dust thermal emission becomes important.  With
higher metallicity, even lower temperature due to the CO cooling makes
OH and H$_2$O hard to be excited and their importance in cooling
diminishes although, for [M/H]=-2, OH cooling is still important for
$10^{6}-10^{7}{\rm cm^{-3}}$.  For [M/H] $\ga -2$, the fine-structure
line transitions of carbon and oxygen are important for cooling in a
diffuse phase ($\la 10^{3} {\rm cm^{-3}}$), which is below our initial
densities and not shown in Figure \ref{fig:cool.all}.  At these
metallicities, CO becomes the dominant coolant before it is overtaken
by dust thermal emission. For [M/H]=0, this shift happens $\sim
10^{3}{\rm cm^{-3}}$ and the CO cooling does not appear in Figure
\ref{fig:cool.all}.

\subsubsection{H$_2$ Formation on Dust Grains}

In a metal-free gas, only a small amount of H$_2$ with concentration
$y({\rm H_2}) \sim 10^{-3}$ is formed via the gas-phase H$^{-}$
reaction channel before the onset of the three-body H$_2$ formation at
$\sim 10^{8}{\rm cm^{-3}}$.  In a metal-enriched gas, dust grains are
usually present as well; efficient H$_2$ formation proceeds on the
grain surface even at low densities.  Figure \ref{fig:nH2} shows that,
at metallicity [M/H] $\ga -5$, the H$_2$ concentration continues
increasing owing to the formation reaction on the dust, unlike the
metal-free case where there is a plateau at $\sim 10^{-3}$. For higher
metallicities, i.e., more dust grains, the H$_2$ formation is more
efficient, and its concentration is higher at the same density before
becoming fully molecular by the three-body reaction ($\la 10^{10}{\rm
cm^{-3}}$).  In the case of [M/H]=-5, for example, about three times
more H$_2$ are formed by the beginning of three-body reaction
($10^{8-9}{\rm cm^{-3}}$) than in the metal-free case.  This higher
H$_2$ abundance causes a slight deviation in temperature from the
metal-free case $\la 10^{7}{\rm cm^{-3}}$ even before the small
temperature decrease by the H$_2$O cooling in the range $10^{7-10}{\rm
cm^{-3}}$ (Fig. \ref{fig:nTc.all}).  With higher metallicity of
[M/H]=-(3-4), further H$_2$ enhancement makes the temperature fall
below the threshold for HD formation at $\simeq 150$K.  The
temperature decreases by HD line cooling until its rotational levels
reach LTE at $\sim 10^{5}{\rm cm^{-3}}$ (Fig. \ref{fig:nTc.all}). With
even higher metallicity ([M/H] $\ga$ -2), despite of higher H$_2$
abundance, the line cooling by metalic atoms (OI) and molecules (CO,
OH) is more significant than that with H$_2$ and HD molecules.

Except for the case of the solar metallicity ([M/H]=0), H$_2$
formation has not yet been completed by the initial density of
calculation at $10^{4}{\rm cm^{-3}}$.  Heating associated with H$_2$
formation has large effects on the temperature evolution.  In the
cases with lowest metallicities ([M/H] $\leq$ -5), the heating by
three-body reaction is dominant over the compressional heating in the
range $10^{9}-10^{12}{\rm cm^{-3}}$. For metallicity of [M/H]=-4..-1,
because of the low temperature at the onset of rapid H$_2$ formation,
the chemical heating develops a remarkable bump in the temperature
evolution, separating two temperature dips, i.e. one at lower density
by line cooling and the other at higher density by dust cooling.  In
particular, this effect makes a large jump in temperature evolution at
$10^{8} {\rm cm^{-3}}$ for [M/H]=-4..-3.

It should be noted that the rapid H$_2$ formation phase and thus the
resultant temperature bump shift towards lower density with higher
metallicity.  Also, the local maximum value of the temperature
decreases with increasing metallicity.  This is because, for the H$_2$
formation in lower density, a larger fraction of the chemical binding
energy is radiated away before collisions with other gas particles
make it available for heating, along with more efficient radiative
cooling by metals.

\subsubsection{Thermal Emission of Dust Grains}

Thermal energy of gas is efficiently transferred via collisions to
dust grains, which has a lower temperature.  The excess energy
deposited on the grain is readily radiated as thermal radiation.  This
works as an efficient cooling mechanism at high densities where the
collisions between gas and dust are frequent enough (see panels in
Fig. \ref{fig:cool.all}), and makes a temperature dip for [M/H] $\ga -
5$ (Fig. \ref{fig:nTc.all}).  The frequent collisions equalize the
temperature of gas and dust (so-called dust-gas coupling). However,
before the density increases enough that the dust-gas coupling occurs,
the dust is cooler than the gas by a large margin. Note that we assume
there is no external background radiation.  In this case, the cooling
rate by dust thermal emission obeys $\Lambda_{\rm dust}=\Gamma_{\rm
coll} \propto Z n_{\rm H} T^{3/2}$ (see eq. \ref{eq:Gcoll}), where the
$Z$ dependence comes from $(f/\rho)_{\rm gr}$.  The compressional
heating rate is proportional to $n_{\rm H}^{1/2} T$ for the collapse
at a rate of the free-fall.  With increasing density, thus, the dust
cooling becomes more important relative to the compressional heating
and finally catches up with it. At this point the gas temperature
falls down until it reaches the dust temperature. Thereafter, the
cooling rate by dust emission is expressed as $\Lambda_{\rm dust}
\propto Z T^{\beta+4}$ in the optically thin case, where the Planck
mean opacity of dust $\kappa_{\rm d} \propto T^{\beta}$.  With keeping
thermal balance between the dust cooling and compressional heating,
temperature slowly increases as $\propto n_{\rm
H}^{1/2(\beta+3)}\simeq n_{\rm H}^{0.1}$, where $\beta \simeq 2$ is
used in the last expression.
 
Because of the dependence of $\Lambda_{\rm dust}$ on $Z$, the dust
cooling becomes less efficient for the lower metallicity and then the
onset of its efficient cooling is postponed until the higher density
regime, where the temperature is also higher.  With metallicity below
a threshold value ([M/H]=-6..-5), the temperature reaches the dust
evaporation value before the dust cooling becomes efficient. In the
case of [M/H]=-6, for example, the dust completely sublimates at $\sim
1500$K ($2\times 10^{14}{\rm cm^{-3}}$) before its cooling reaching
the level of compressional heating, although it does some contribution
at $10^{13}-10^{14}{\rm cm^{-3}}$.  The temperature dip by dust
cooling is absent in this case.  With higher metallicities, the dust
cooling reaches the compressional heating at some point and balances
it thereafter until the dust thermal emission becomes optically thick
and ineffective.  After the core becomes optically thick to the dust
absorption, the temperature increases almost adiabatically with
increasing density as seen clearly for [M/H] $\geq -4$ in Figure
\ref{fig:nTc.all}.  This phase is called the {\it first adiabatic
phase}. We distinguish it from another adiabatic phase ({\it second
adiabatic phase}) at a higher density ($\ga 10^{21}{\rm cm^{-3}}$).
Note that a cloud core with higher metallicity becomes optically thick
to dust absorption at progressively lower density and temperature: for
example, $10^{11}{\rm cm^{-3}}$ for $Z_{\sun}$, and $10^{13}{\rm
cm^{-3}}$ for $10^{-4}Z_{\sun}$.  Nevertheless, evolutionary paths of
temperatures (Fig. \ref{fig:nTc.all}) during and after the first
adiabatic phase are similar each other for [M/H] $\ga -5$ (see Omukai
2000 for further explanation). Finally, at $\simeq$ 1500K, the
remaining dust species, i.e., silicates and iron, totally evaporate.
The central part of the core is now free of dust and becomes
transparent again although the surrounding matter is still optically
thick.  Because the surrounding gas has a lower temperature, the
central part is still able to cool by continuum emission.  This phase,
however, lasts for a brief interval since the core soon becomes
optically thick to the same continuum.  The H$_2$ dissociation begins
at this moment and the temperature increases only gradually under the
action of the dissociation cooling. After its completion at $\sim
10^{21}{\rm cm^{-3}}$, the temperature increases adiabatically again.
This is the second adiabatic phase.  The evolution after the beginning
of H$_2$ dissociation is similar to the metal-free case, although the
temperature is somewhat lower than that in the metal-free case.

\subsection{Dynamical Evolution}

We next describe dynamical evolution of the dense cores as a response
to the thermal evolution seen above in Sec. 3.2. In Figure
\ref{fig:colrate}, the rate of enhancement of central density
$\rho_{\rm c}$, which can be given by the inverse of the dynamical
timescale, $t_{\rm dyn}^{-1}=(d\rho_{\rm c}/dt)/\rho_{\rm c}$ is shown
relative to the inverse of the free-fall timescale $t_{\rm
ff}^{-1}=1/\sqrt{\frac{3 \pi}{32 G \rho_{\rm c}}}$.  By comparing
Figures \ref{fig:gamma} and \ref{fig:colrate}, we see an
anti-correlation between the collapse rate $t_{\rm ff}/t_{\rm dyn}$
and $\gamma$: where $\gamma$ is high (low), the collapse rate is low
(high) owing to increasing (decreasing, respectively) pressure with
collapse relative to the gravity.  In particular, when $\gamma$
exceeds 4/3, the dynamical collapse is halted and the collapse rate
becomes very low with $t_{\rm ff}/t_{\rm dyn}< 1$ (lower dotted
horizontal lines in Fig. \ref{fig:colrate}), indicating that the
collapse is quasi-static, although the core contracts slowly by
accretion of ambient matter.

\begin{figure}
\plotone{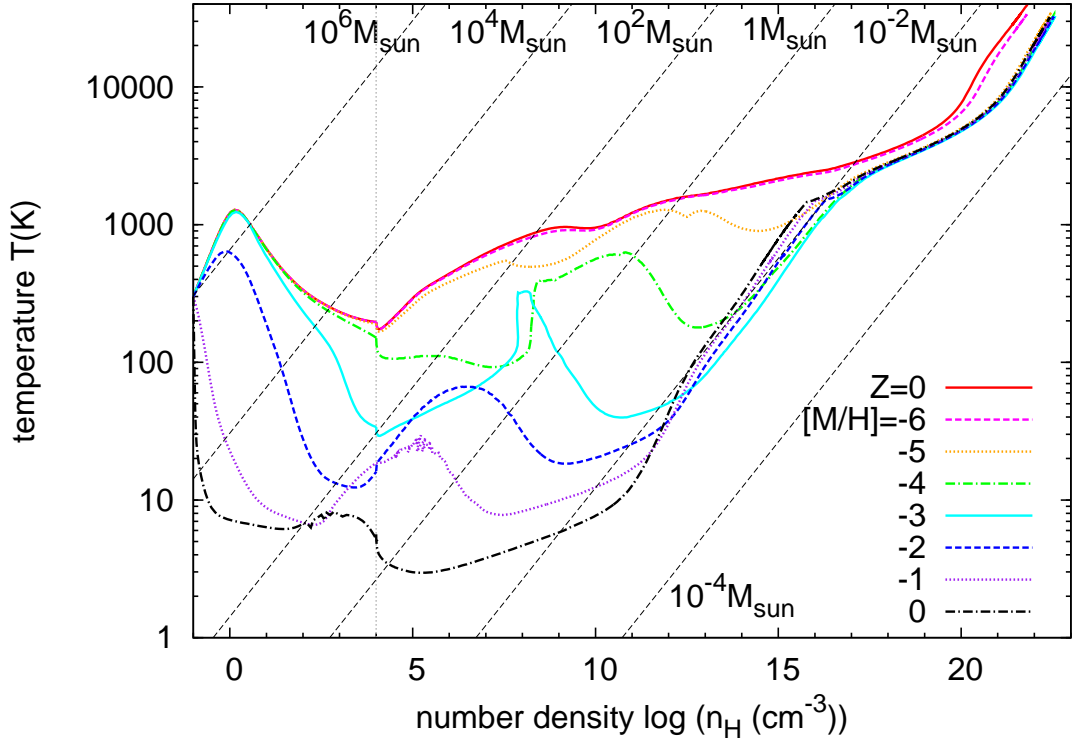} \caption{The evolution of temperatures
at the center of cloud cores during the prestellar collapse for
various metallicities.  This is calculated by one-zone model until
$10^{4}{\rm cm^{-3}}$ (dotted vertical line) and by hydrodynamical
models for the higher density.  The constant Jeans masses are
indicated by the dashed lines.
\label{fig:nTc.all}}
\end{figure}

\begin{figure}
\plotone{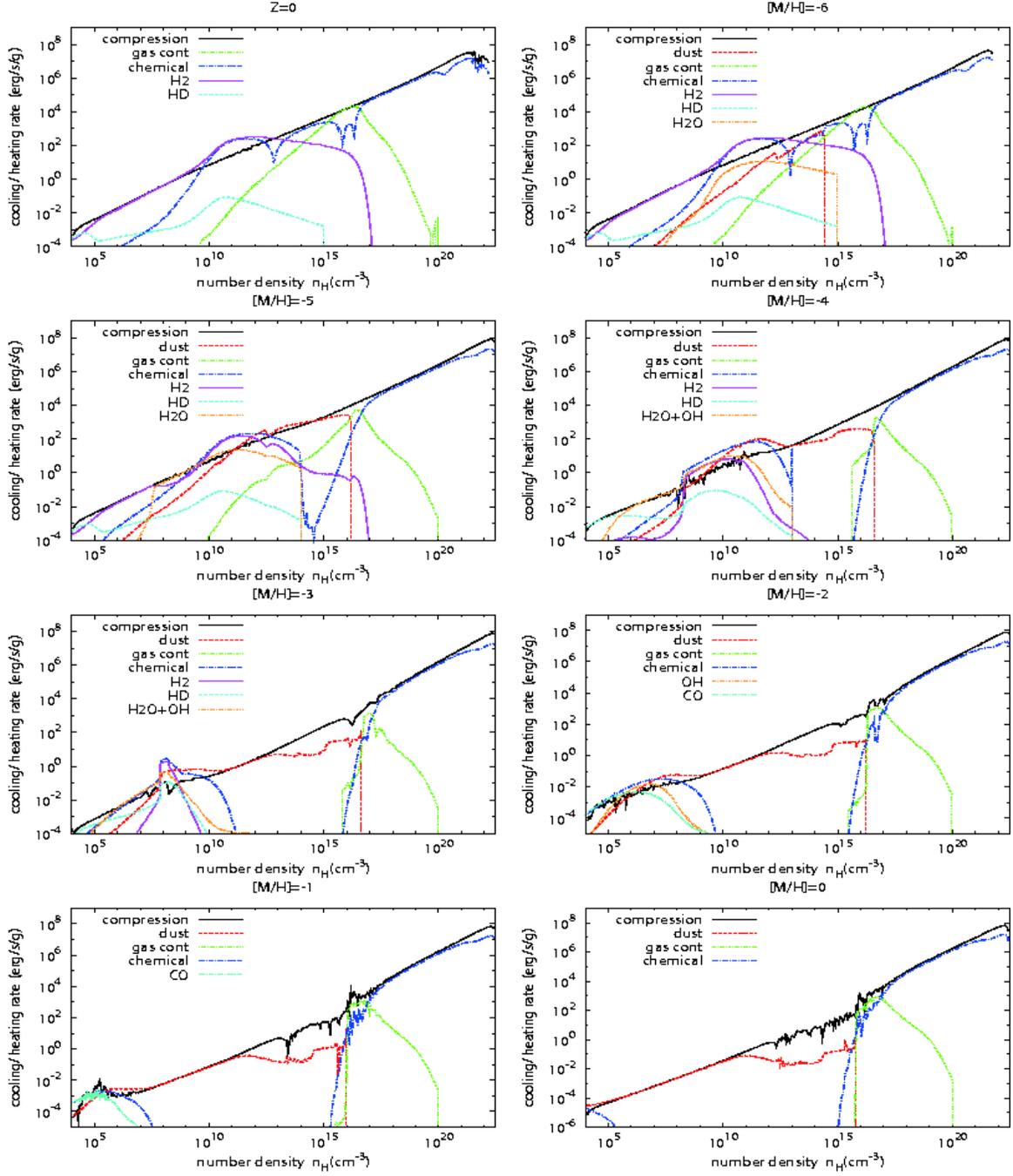} \caption{The cooling and heating rates
per unit mass by individual processes at the centers of the cloud
cores as functions of number density.  To avoid violent oscillatory
behaviors and keep the compressional heating rate positive during the
adiabatic phases, we plot them at every time when the density exceeds
1.1 times the previous values.
\label{fig:cool.all}}
\end{figure}

\begin{figure}
\plotone{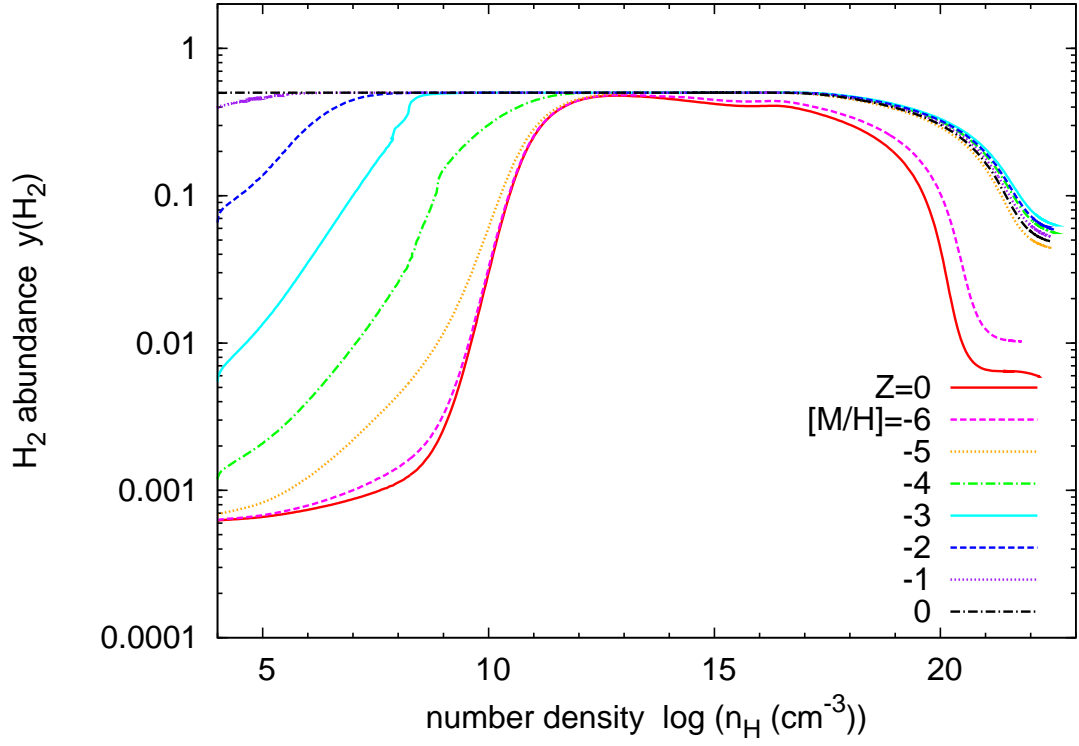} \caption{The evolution of the H$_2$
concentrations $y({\rm H_2})$ at the centers of the cores. The state
of $y({\rm H_2})=1/2$ corresponds to the fully molecular gas.
\label{fig:nH2}}
\end{figure}

\begin{figure}
\plotone{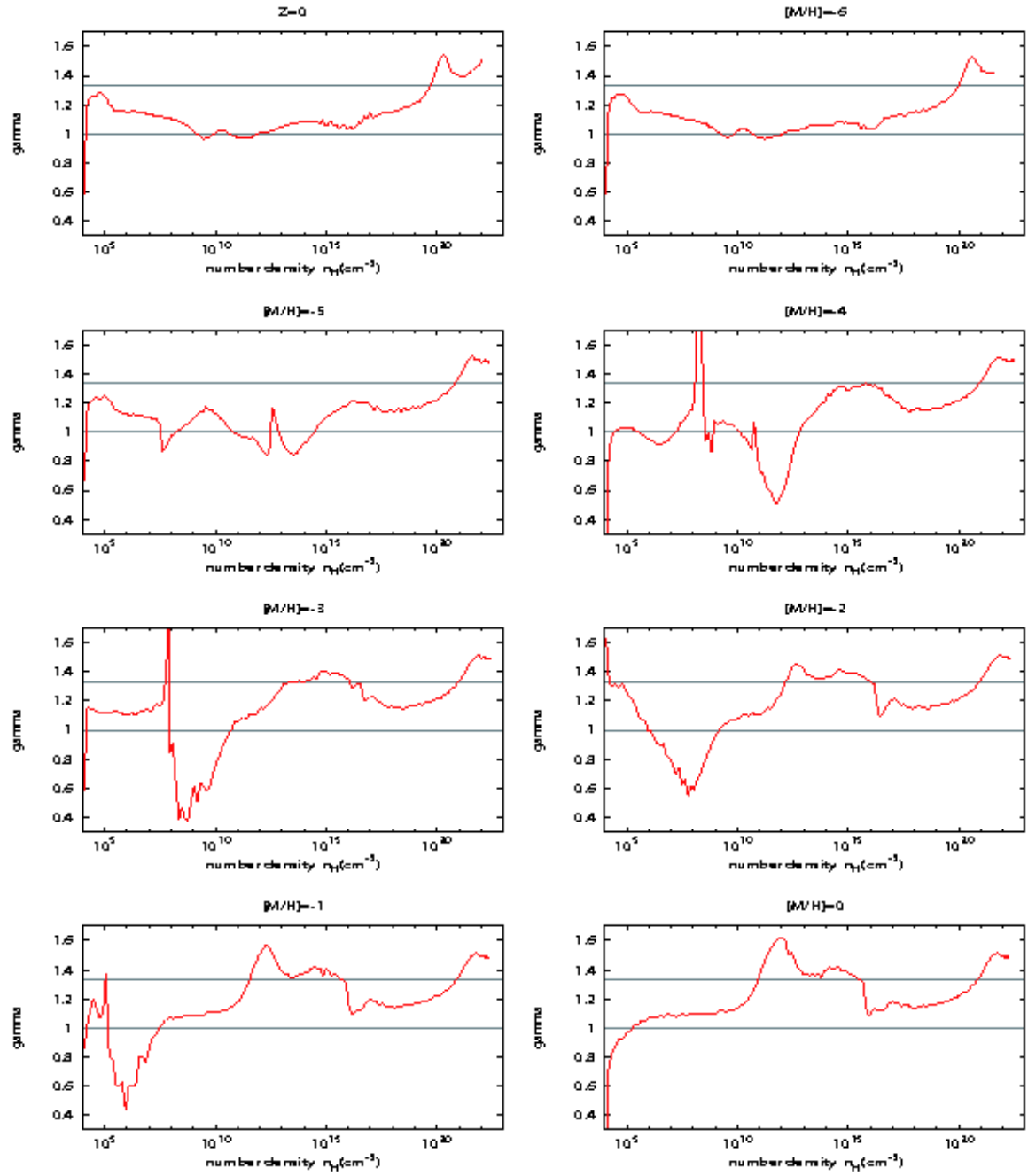} \caption{The effective ratio of
specific heat $\gamma={\rm dlog} p/{\rm dlog} \rho$ at the center
during the prestellar collapse for different metallicities as a
function of number density.  The horizontal lines show the critical
values for fragmentation (1) and for hydrostatic core formation (4/3),
respectively.
\label{fig:gamma}}
\end{figure}

\begin{figure}
\plotone{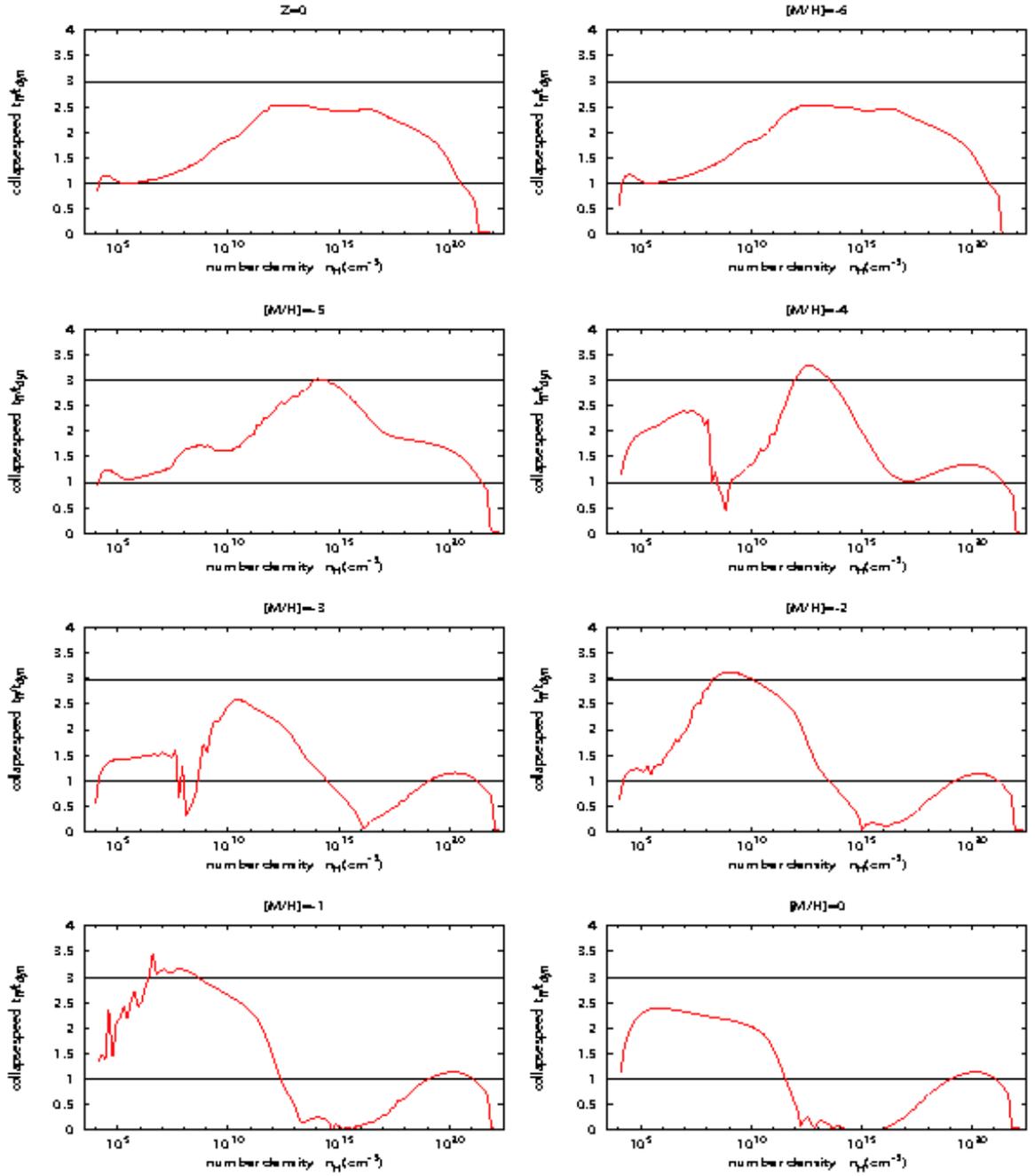} \caption{ The collapse rate $t_{\rm
dyn}^{-1}=\dot{\rho}/\rho$ at the center, normalized by the free-fall
rate $t_{\rm ff}^{-1}$.  The dashed horizontal curves indicate the
value for the isothermal Larson-Penston similarity solution (3.0) and
the threshold for quasi-static collapse (1.0).
\label{fig:colrate}}
\end{figure}

\begin{figure}
\plotone{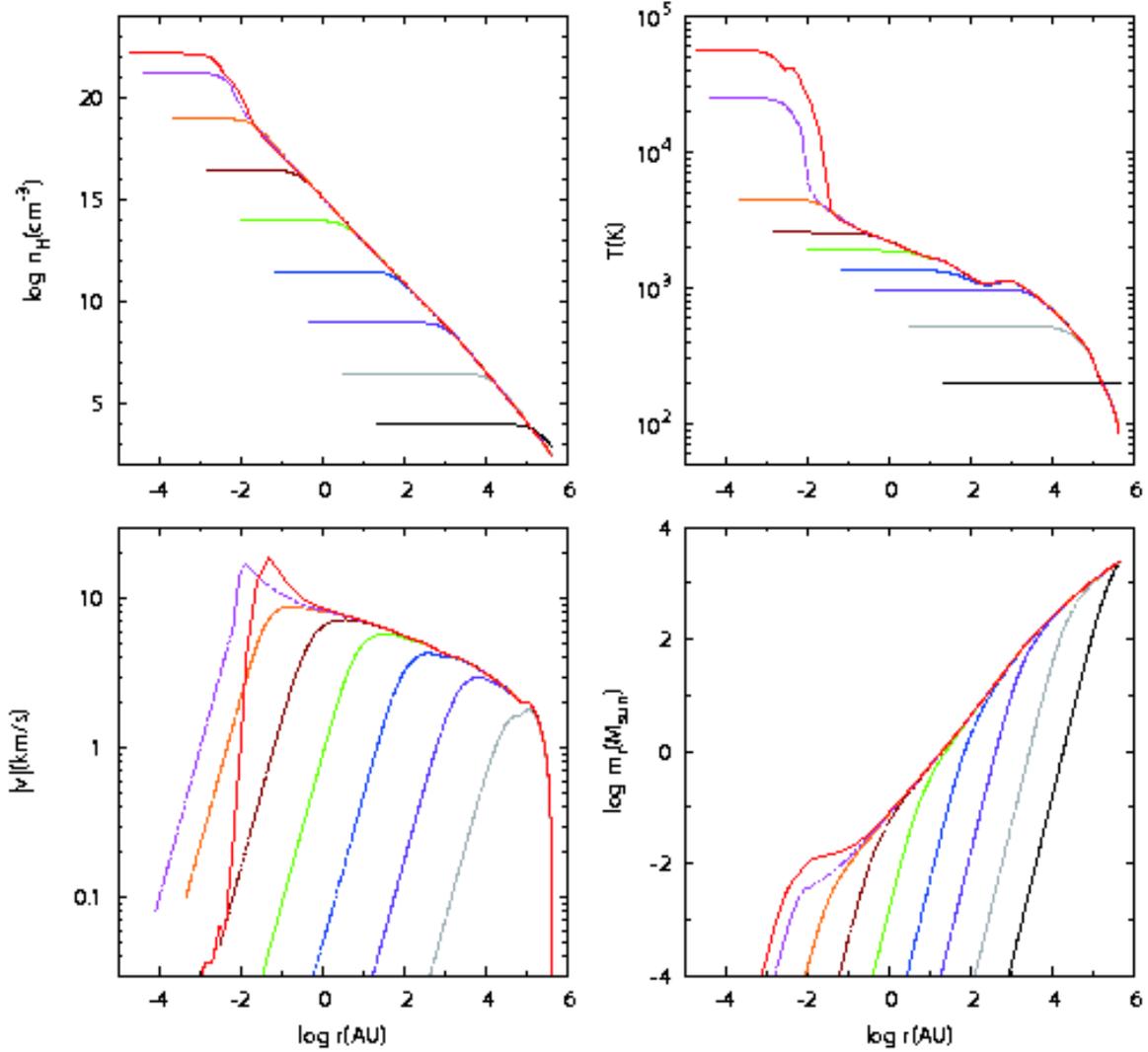} \caption{The radial distributions of
(a) density (b) temperature (c) velocity and (d) enclosed mass at
different epochs during the prestellar collapse for the cloud core
with metallicity Z=0.  The first curves shown correspond to the
initial state of the calculation.  They are plotted when the central
density is enhanced by a factor of $10^{2.5}$ from the previous one
except the last state shown, which corresponds to the time somewhat
after the formation of a protostar.
\label{fig:four.zero}}
\end{figure}

\begin{figure}
\plotone{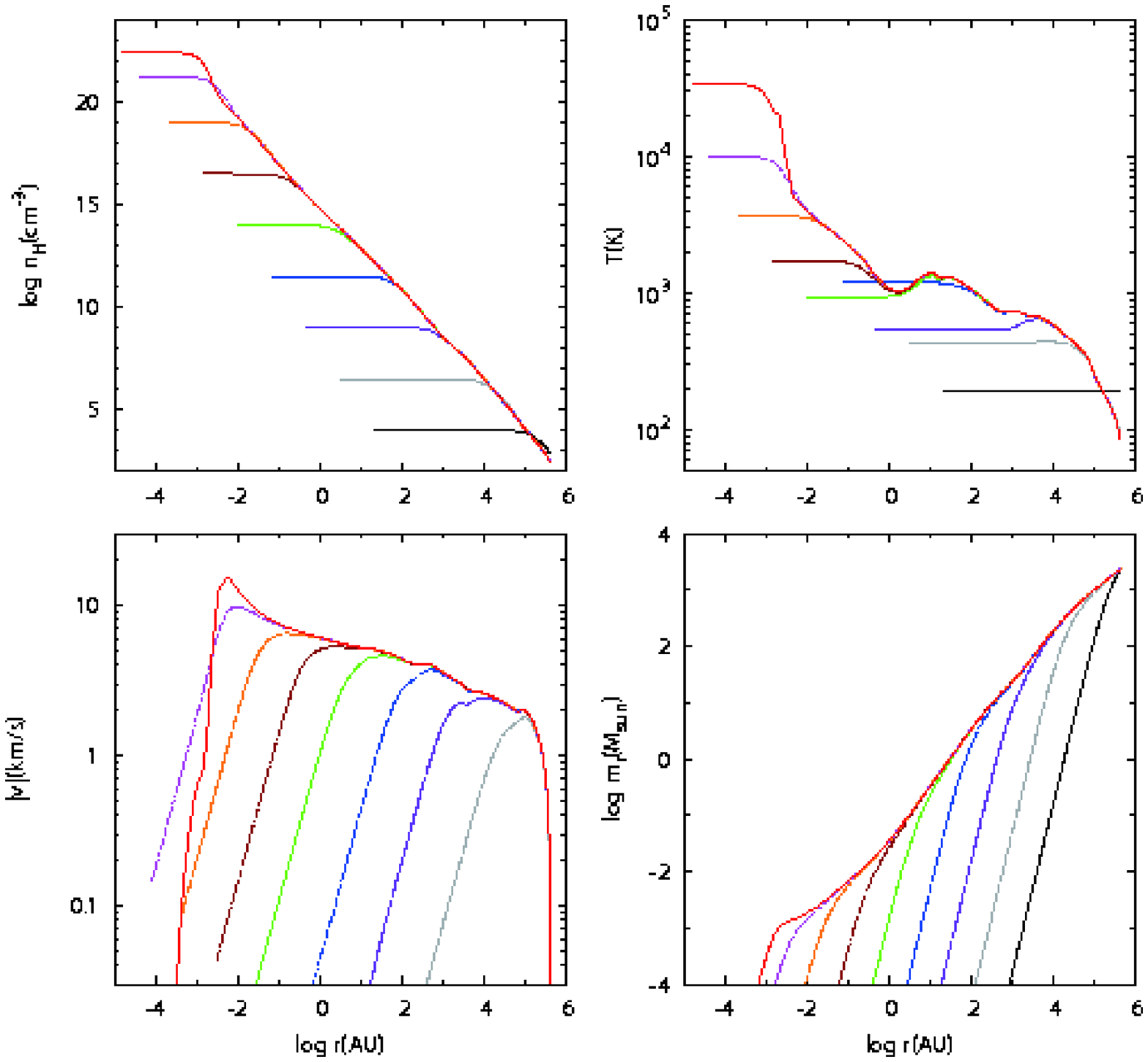} \caption{The same as Figure
\ref{fig:four.zero}, but for metallicity [M/H]=-5
\label{fig:four.-5}}
\end{figure}

\begin{figure}
\plotone{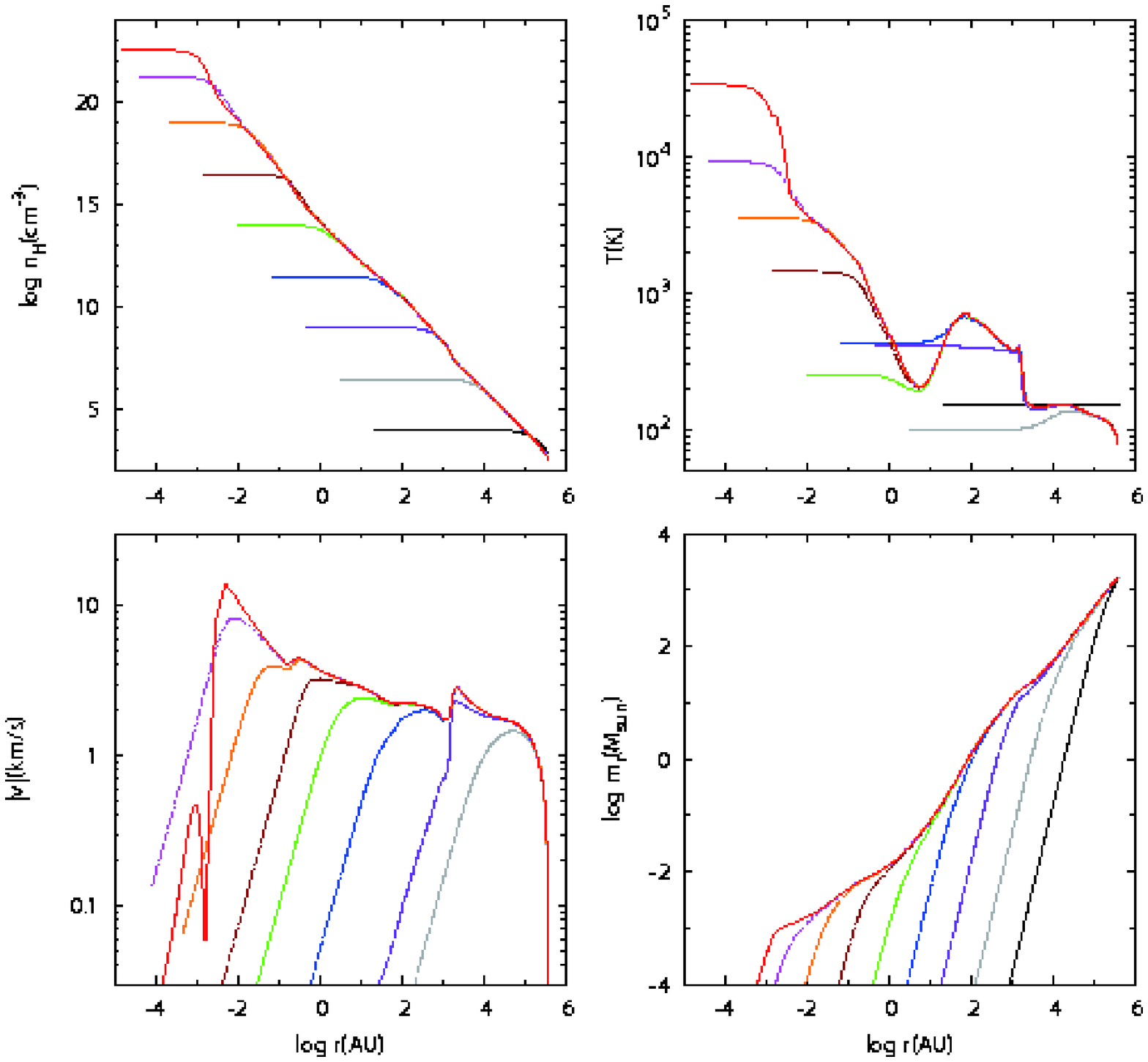} \caption{The same as Figure
\ref{fig:four.zero}, but for metallicity [M/H]=-4
\label{fig:four.-4}}
\end{figure}

\begin{figure}
\plotone{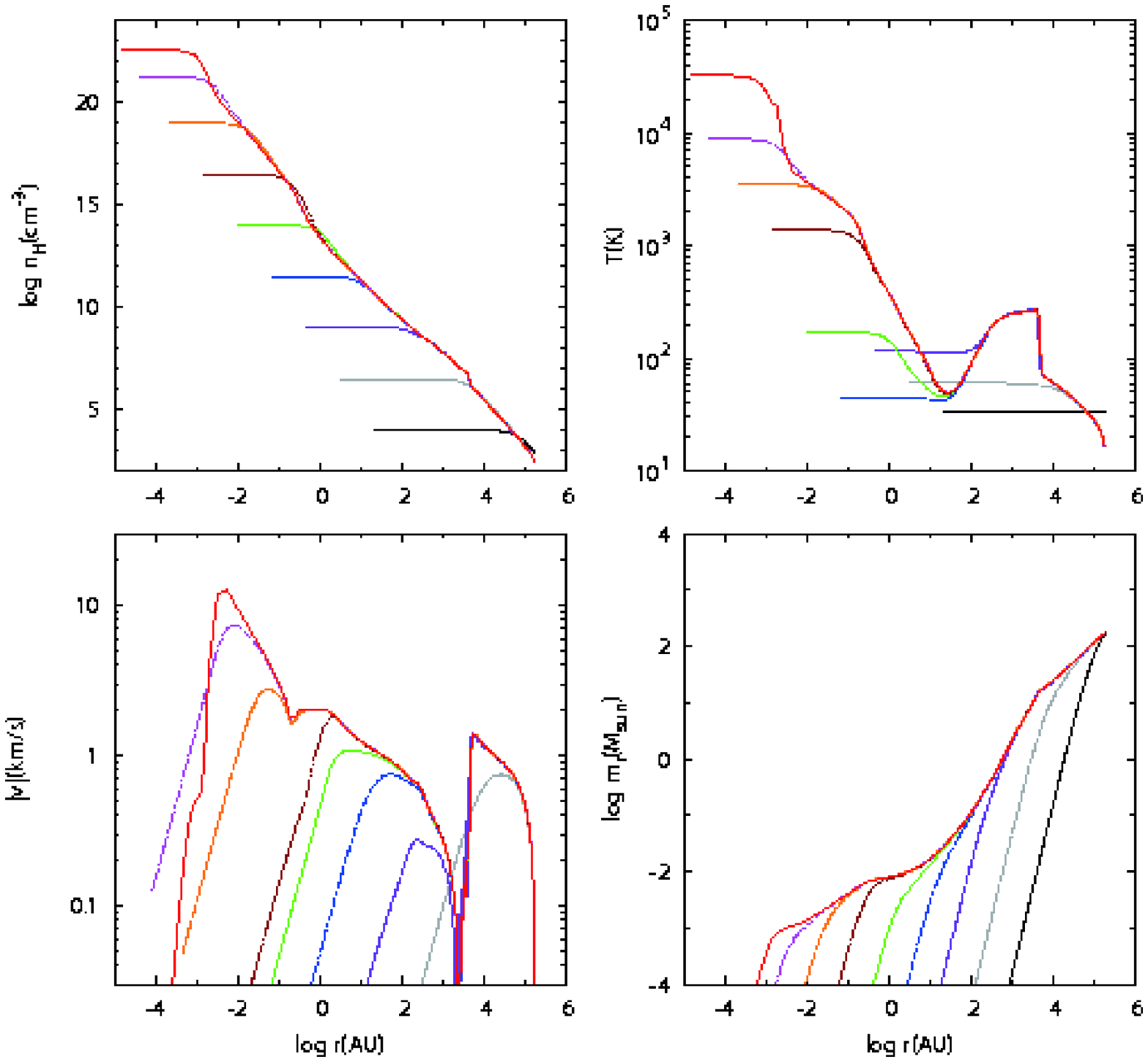} \caption{The same as Figure
\ref{fig:four.zero}, but for metallicity [M/H]=-3
\label{fig:four.-3}}
\end{figure}

\begin{figure}
\plotone{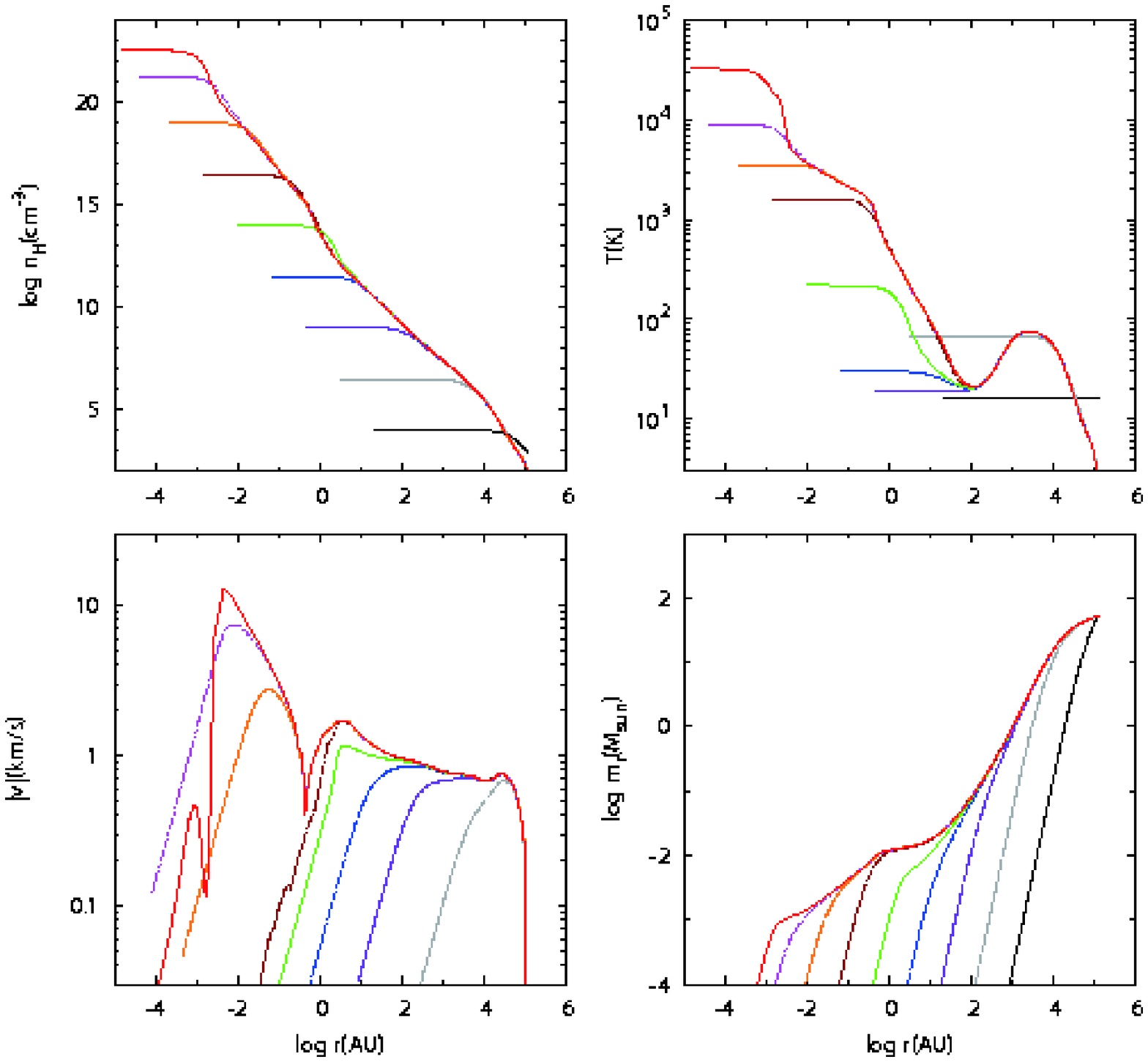} \caption{The same as Figure
\ref{fig:four.zero}, but for metallicity [M/H]=-2
\label{fig:four.-2}}
\end{figure}

\begin{figure}
\plotone{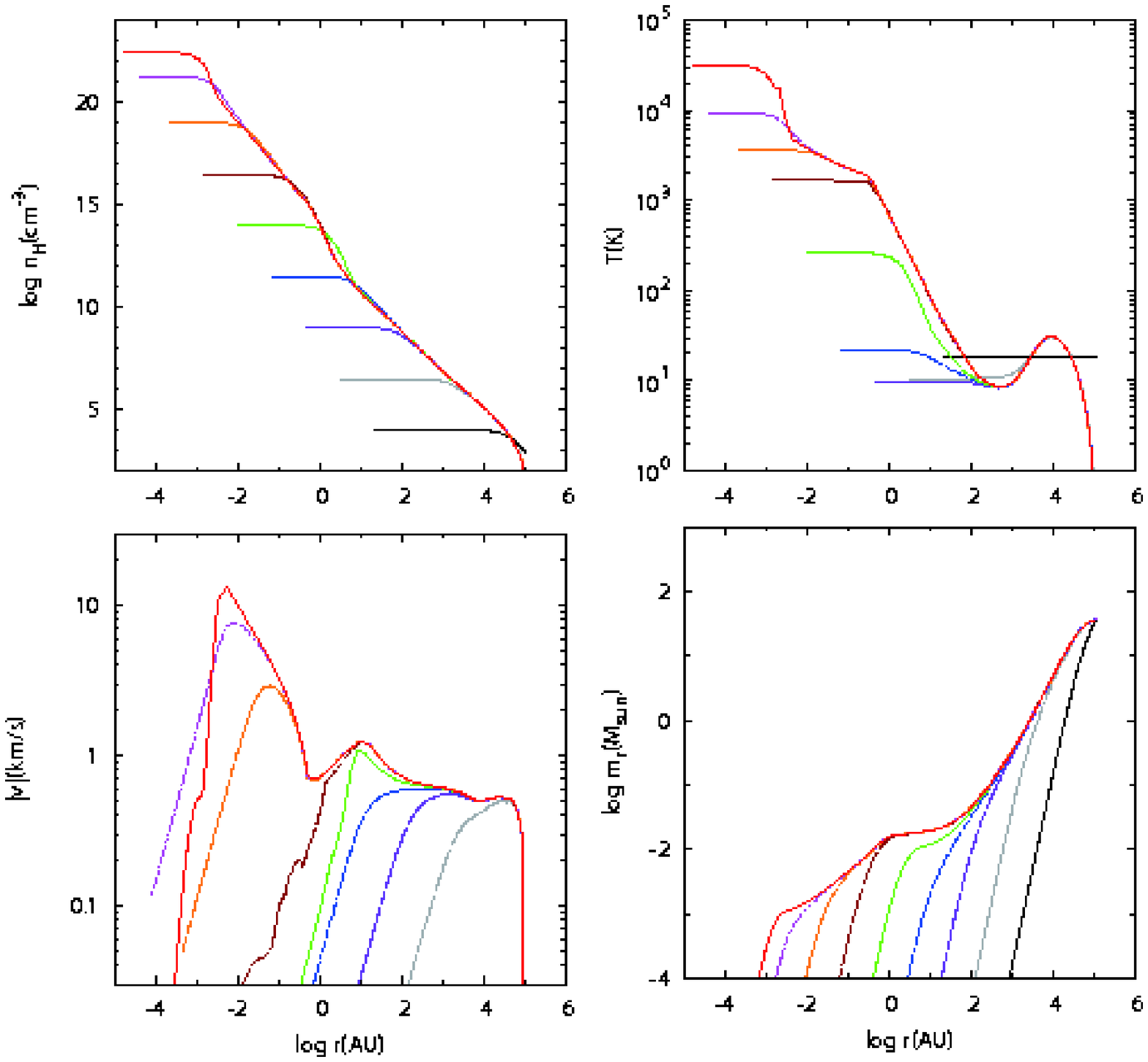} \caption{The same as Figure
\ref{fig:four.zero}, but for metallicity [M/H]=-1
\label{fig:four.-1}}
\end{figure}

\begin{figure}
\plotone{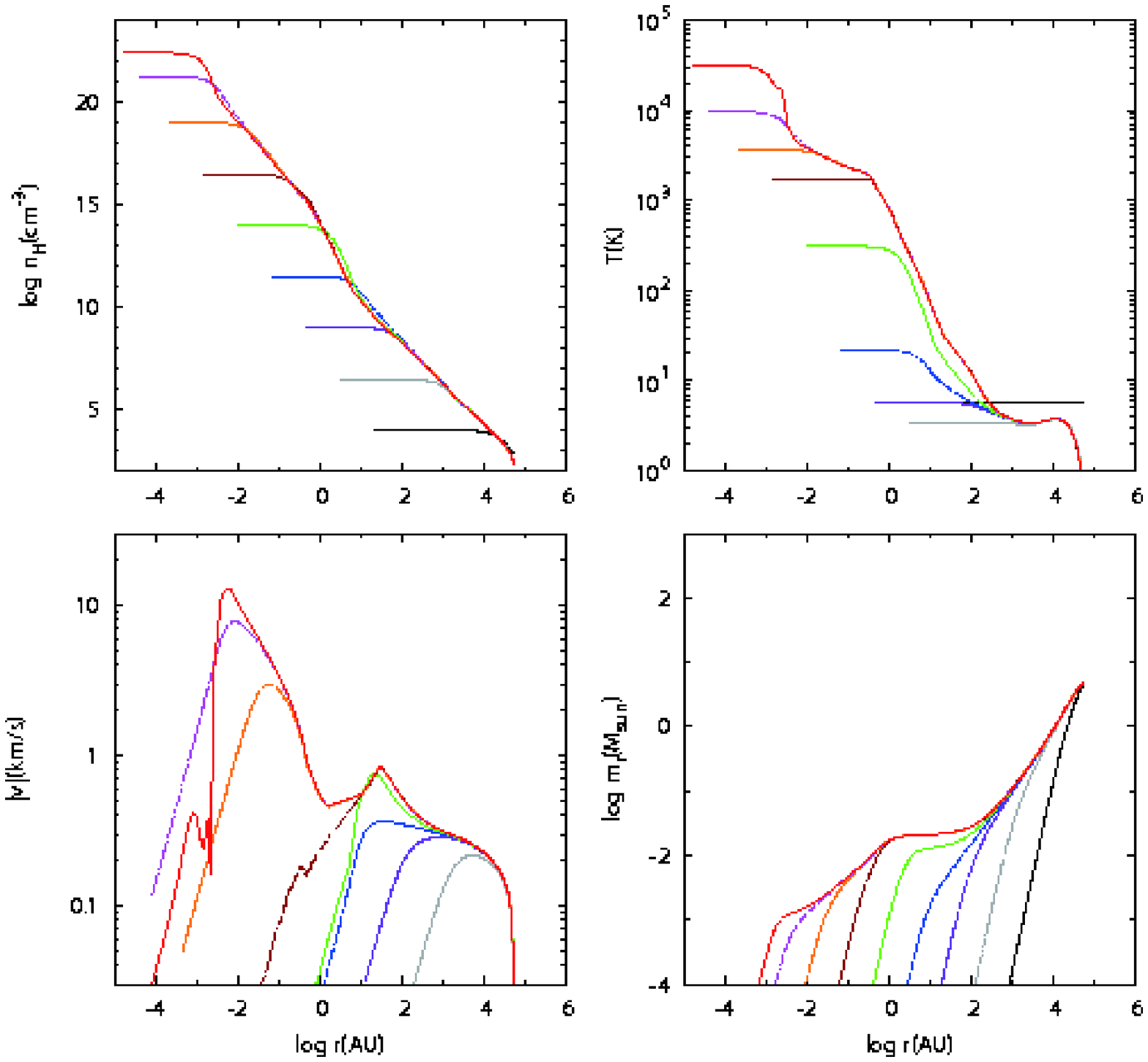} \caption{The same as Figure
\ref{fig:four.zero}, but for metallicity [M/H]=0
\label{fig:four.sun}}
\end{figure}

In Figures \ref{fig:four.zero}-\ref{fig:four.sun}, we present radial
distributions of density, temperature, velocity, and enclosed mass of
the dense cores at different epochs until the formation of protostars
for zero-metallicity, and metallicity of [M/H]=-6, -5, -4, -3, -2, -1,
and 0.  Between successive plots, the increment of central density is
$10^{2.5}$ times that of the previous one.  The exception is the final
phase shown, which corresponds to the state slightly after the end of
the prestellar collapse. Similarity nature of the collapse is obvious
in density distributions in Panels (a) of Figures
\ref{fig:four.zero}-\ref{fig:four.sun}.  This collapse closely
resembles that of the Larson-Penston (LP) similarity solution (Larson
1969; Penston 1969), which describes the runaway collapse of
isothermal clouds. In this solution, a flat-density central part of
about a Jeans length collapses at a rate of $t_{\rm ff}/t_{\rm
dyn}=3.0$. With increasing density, the Jeans mass and thus the matter
in the central flat region decreases.  Since the collapse timescale
$t_{\rm dyn} \propto \rho_{\rm c}^{-1/2}$ becomes shorter and shorter
with increasing density, a gas in low-density envelope is effectively
leftover from the evolution of the central region during the collapse.
The density distribution in the envelope has a power of index -2 with
4.4 times that of the singular isothermal sphere, while the infall
velocity is constant at 3.3 times the sound speed in the LP solution.
The same type of solutions for more general polytropic equation of
state has been found by Yahil (1980).
Figure \ref{fig:colrate} shows that the collapse rate at the center of
our cores also becomes close to the value for the LP solution, which
is $t_{\rm ff}/t_{\rm dyn}=3.0$ for the isothermal case and indicated
by dotted horizontal line.  Note that the value $t_{\rm ff}/t_{\rm
dyn}$ higher than unity does not mean the collapse is faster than the
free-fall, for which $t_{\rm ff}/t_{\rm dyn}=3\pi/2=4.7$
asymptotically.  In some occasions, in particular for the cores in the
dust-cooling phases, peak values of the collapse rate exceed the
isothermal LP value of 3.0 slightly as $\gamma< 1$ in those
ranges. For [M/H]=-4 and -3, the temperature jump by H$_2$ formation
heating at $\sim 10^{8}~{\rm cm^{-3}}$ is observed as a sharp peak in
$\gamma$ in Figure \ref{fig:gamma}. The slow-down of the collapse at
this epoch results in a temporary small value of $t_{\rm ff}/t_{\rm
dyn} \sim 0.5$ (Fig. \ref{fig:colrate} as downward spikes), which
indicates that the collapse becomes quasi-static.  Although the sudden
slow-down of the collapse develops a shock around the heated core (see
Figs. \ref{fig:four.-4} and \ref{fig:four.-3}c), this heating phase
lasts only for a brief period and then dynamical collapse is soon
recovered.  While H$_2$ formation is in progress, small oscillations
in the central density are imposed by the associated heating on the
overall collapse.  These oscillations are seen for [M/H]=-1 as notches
around the peak of the bump in Figure \ref{fig:nTc.all}.  For other
cases, they are not noticeable in Figure \ref{fig:nTc.all}, but still
observable in Figure \ref{fig:cool.all} as notches in the
compressional heating rates at H$_2$ formation epoch.
After the core becomes optically thick, temperature increases
adiabatically and $\gamma$ exceeds 4/3.  The pressure overcomes the
gravitational pull before long and a hydrostatic core forms.  This
hydrostatic core is often referred to the {\it first protostellar
core} in literature as it appears before the genuine protostar, which
is also called the {\it second protostellar core}.  Although dynamical
collapse is halted in this phase, the core continues contracting on
average, but its oscillates acoustically many times, in response to
the increasing mass of the hydrostatic core by accretion of the
envelope material. Finally, at $\sim 10^{16}{\rm cm^{-3}}$, dynamical
collapse is resumed owing to the cooling by H$_2$ dissociation.  When
the dissociation has almost completed, the temperature again begins to
increase adiabatically.  A hydrostatic core formed in this second
adiabatic-phase grows eventually to a star and thus is called the
protostar.  The formation of the protostar marks the transition from
the prestellar phase to the protostellar phase. Observationally, the
protostars still in the early protostar phase are considered be the
Class 0 objects, while the Class I objects are later but still in the
protostar phase.

In the metal-free case, the temperature increases gradually at almost
a constant rate from $10^{4}~{\rm cm^{-3}}$ to $10^{19}~{\rm cm^{-3}}$
without remarkable features.  Consequently, the radial density profile
does not deviate noticeably from that of the LP-type similarity
solution until $\ga 10^{20}~{\rm cm^{-3}}$, where the appearance of a
hydrostatic core results in steepening of density gradient.  For a
polytropic gas whose equation of state is $P\propto \rho^{\gamma}$,
the envelope density profile obeys $\rho \propto r^{-2/(2-\gamma)}$
(Larson 1969; Yahil 1983). Since $\gamma$ takes the values of $1 -
1.2$ at $< 10^{19}~{\rm cm^{-3}}$ (Fig. \ref{fig:gamma}), the density
gradient in the envelope is $\simeq -2.2$, which is slightly steeper
than the value for the isothermal collapse, $-2$.  The three-body
H$_2$ formation makes a small dip in temperature evolution at $\sim
10^{10}{\rm cm^{-3}}$ (Fig. \ref{fig:nTc.all}), which is observed also
in radial temperature distribution at $10^{2-3}{\rm AU}$
(Fig. \ref{fig:four.zero} b).  The formation of the protostar at
$10^{21-22}{\rm cm^{-3}}$ results in sudden halting of the inflow,
which develops the accretion shock at the stellar surface $\sim
10^{-2}{\rm AU}$ (Fig. \ref{fig:four.zero} c). Also, the protostar
formation can be seen as steepening of density profile around the
protostellar surface, rapid increase of temperature, and a plateau in
the enclosed mass profile. The mass of the protostar at its birth is
somewhat smaller than $\sim 10^{-2}M_{\sun}$.
With metal enrichment as small as [M/H]=-5, the density, velocity, and
mass profiles look similar to those in the metal-free case, although
temperature profile deviates from the metal-free one at $\sim
10^{3}{\rm AU}$ due to H$_2$O cooling and $\sim 1{\rm AU}$ due to dust
cooling (Fig. \ref{fig:four.-5}).  Owing to the difference in
temperature evolution for $\ga 10^{16} {\rm cm^{-3}}$
(Fig. \ref{fig:nTc.all}), the effective adiabatic index $\gamma$
exceeds the critical value 4/3 at $10^{21} {\rm cm^{-3}}$ for [M/H]=-5
and higher metallicity cases, while at $10^{20} {\rm cm^{-3}}$ for
$Z=0$ (see Fig. \ref{fig:gamma}).  Thus the formation of protostar is
delayed to slightly higher density and, the mass and radius at
protostar formation are somewhat smaller than the metal-free case.

For the case of [M/H]=-4, the heating by H$_2$ formation makes a jump
in temperature distribution at $\sim 10^{3} {\rm AU}$
(Fig. \ref{fig:four.-4}b). This causes slow-down of collapse and
formation of a short-lived hydrostatic core, whose appearance can be
seen as kinks in the density and mass profiles
(Figs. \ref{fig:four.-4}a, d), and a shock in the velocity profile
(Fig. \ref{fig:four.-4}c). A temperature dip by dust cooling is
obvious at $\sim$ 10AU.  Almost simultaneously with the dust-gas
coupling, the core becomes optically thick to the dust opacity.
Subsequently, the temperature increases rapidly for
$10^{14}-10^{16}{\rm cm^{-3}}$, but $\gamma$ does not exceed 4/3 (Fig. 
\ref{fig:gamma}). The dynamical collapse continues although slowing
down with $t_{\rm ff}/t_{\rm dyn}$ reaching unity at $10^{17}{\rm
cm^{-3}}$, which indicates the collapse is now quasi-static
(Fig. \ref{fig:colrate}).  This slow-down is seen as a small spike at
$\sim$ 0.1 AU in the velocity profile (Fig. \ref{fig:four.-4}c).  Even
in a non-spherical collapse, a cloud becomes round by this heating and
fragmentation in the later dust-cooling phase might be suppressed.  If
so, low-mass fragments and stars cannot be formed in the range [M/H]
$\sim -4$.  Tsuribe \& Omukai (2008) attributed the origin of apparent
paucity of stars in the range [M/H] = $-(4.8..4.0)$ in the halo to it. 
The evolution after the H$_2$ dissociation $\ga 10^{16}{\rm cm^{-3}}$
is identical to the case of [M/H]=-5.

The case of [M/H]=-3 is qualitatively similar to [M/H]=-4
(Fig. \ref{fig:four.-3}). However, the H$_2$ formation heating, as
well as the temperature increase in the optically thick regime, have
larger effects. Even before the rapid H$_2$-formation heating at
$10^{8}{\rm cm^{-3}}$, slower heating has started and temperature
increases gradually with $\gamma=1.1-1.2$, higher than $\simeq 1$ for
[M/H]=-4 in the same regime. The collapse then is slower with $t_{\rm
ff}/t_{\rm dyn} \simeq 1.5$ while it is $\simeq 2$ for [M/H]=-4.  Thus
the consequence of the rapid heating is almost complete halt of the
collapse for a brief period, as observed in a deep trough in the
velocity profile at 2000AU (Fig. \ref{fig:four.-3}c).  The central
part becomes optically thick to the dust absorption at $10^{12} {\rm
cm^{-3}}$, and the adiabatic contraction phase continues until
$10^{16-17}{\rm cm^{-3}}$, where the dust sublimates and then the
H$_2$ dissociation follows.  Note that since the adiabatic phase
begins at lower density for higher metallicity and ceases at almost
the same density, the adiabatic phase is more prolonged.  As seen in
Figure \ref{fig:colrate}, the collapse becomes quasi-static for $\ga
10^{15}{\rm cm^{-3}}$ and a hydrostatic object -- the first
protostellar core -- forms.  Signs of the hydrostatic core formation,
i.e., density steepening, temperature increase, velocity dip, and
flattening of enclosed mass profile, are seen at $\sim$ 1 AU in Figure
\ref{fig:four.-3}.

In the case of [M/H]=-2 (Fig. \ref{fig:four.-2}), the H$_2$ formation
heating is gradual with $\gamma \simeq 1.3$ without a jump in
temperature.  Thus, no hydrostatic object forms in this phase.  The
hydrostatic objects (first hydrostatic core) forms at $10^{14} {\rm
cm^{-4}}$, after the central part becomes optically thick to dust
absorption. The core starts collapsing dynamically again by dust
evaporation and H$_2$ dissociation. Finally, at $10^{22}{\rm
cm^{-3}}$, the second hydrostatic core, i.e. protostar, forms with the
same condition as in other cases with [M/H]$\geq -5$.
  
For higher metallicities (Figs. \ref{fig:four.-1} and
\ref{fig:four.sun}), the qualitative nature of the evolution remains
the same, but the onsets of both the efficient dust-cooling and
adiabatic phases are shifted to lower densities. Because of the longer
adiabatic phase, the appearance of the first hydrostatic cores become
more conspicuous (at $\sim$ 1AU). The radius of the first protostellar
core becomes slightly larger with increasing metallicity.  Note that
for the solar metallicity ([M/H]=0) case, the initial (at $10^{4} {\rm
cm^{-3}}$) temperature is already low, the outer bump is missing in
the temperature profile (Fig. \ref{fig:four.sun}b).

\subsection{Comparison with the One-zone Model}

Figures \ref{fig:one-zone} and \ref{fig:nTc.all} show some differences
between the hydrodynamical and one-zone models, despite the overall
similarity.  In general, temperatures tend to be higher in the
hydrodynamical models than in the one-zone models.  The differences
are summarized as follows;
(1) In the one-zone model, all the evolutionary tracks converge for
$\ga 10^{16}{\rm cm^{-3}}$, while in the hydrodynamical models, the
two lowest metallicity cases ($Z=0$ and [M/H]=-6) remain higher than
in others.
(2) At [M/H]=-5, the temperature dip by dust cooling around
$10^{14}{\rm cm^{-3}}$ in hydrodynamical model is not as obvious as in
the one-zone model.
(3) At [M/H]=-3, H$_2$ formation heating results in a large
temperature jump at $\sim 10^{8}{\rm cm^{-3}}$ in the hydrodynamical
model, while in the one-zone model it ends up with a gradual increase
in temperature.

All these deviations can be attributed to the fact that the collapse
rate of the core tends to be faster in the hydrodynamical model than
that assumed in the one-zone model.  In the one-zone model, the
dynamical timescale is assumed to be equal to the free-fall time,
i.e., $t_{\rm ff}/t_{\rm dyn} \simeq 1$.  In the hydrodynamical
models, soon after the beginning of collapse from the rest at
$10^{4}{\rm cm^{-3}}$, the collapse rate exceeds $t_{\rm ff}/t_{\rm
dyn}>1$, which indicates that the collapse is highly dynamical. It
should be noted that the free-fall timescale is the time needed for
the density of the homogeneous gas at rest to reach infinite if the
pressure is negligible, and $t_{\rm ff}/t_{\rm dyn}$ higher than unity
does not mean that the collapse is indeed faster than the free fall.

Observing convergence of temperature tracks in high densities in the
one-zone model, Omukai (2000) demonstrated analytically that the cores
that cool by continuum emission and collapse over a timescale
proportional to the free-fall time have the same specific entropy when
they become optically thick to the same continuum, regardless of their
different metallicities.  The same value of specific entropy assures
the convergence of the evolutionary tracks after the cores become
optically thick. In the discussion of Omukai (2000), however, the gas
was assumed to be fully molecular when the core becomes optically
thick to the continuum. In our hydrodynamical models, due to higher
temperature than in the one-zone models for Z=0 and [M/H]=-6, about
10\% of hydrogen molecules are already dissociated when the core
becomes optically thick to the H$_2$ collision-induced absorption.
The lower binding energy and thus higher specific entropy result in
higher temperatures even after the core becomes optically thick.
 
\section{Evolution of protostars in the accretion phase}
In Sec. 3, we have seen the collapse of dense cores and formation of
tiny protostars at the center.  Although very small mass of $\la
10^{-2}M_{\sun}$ at its birth, the protostar is surrounded by a large
amount of gas in the envelope, which is a remnant of the parental
core.  The protostar thus grows in mass by orders of magnitude through
accretion of these materials.  This is the main accretion phase,
corresponding to Class 0 and I phases, observationally.

\begin{figure}
\plotone{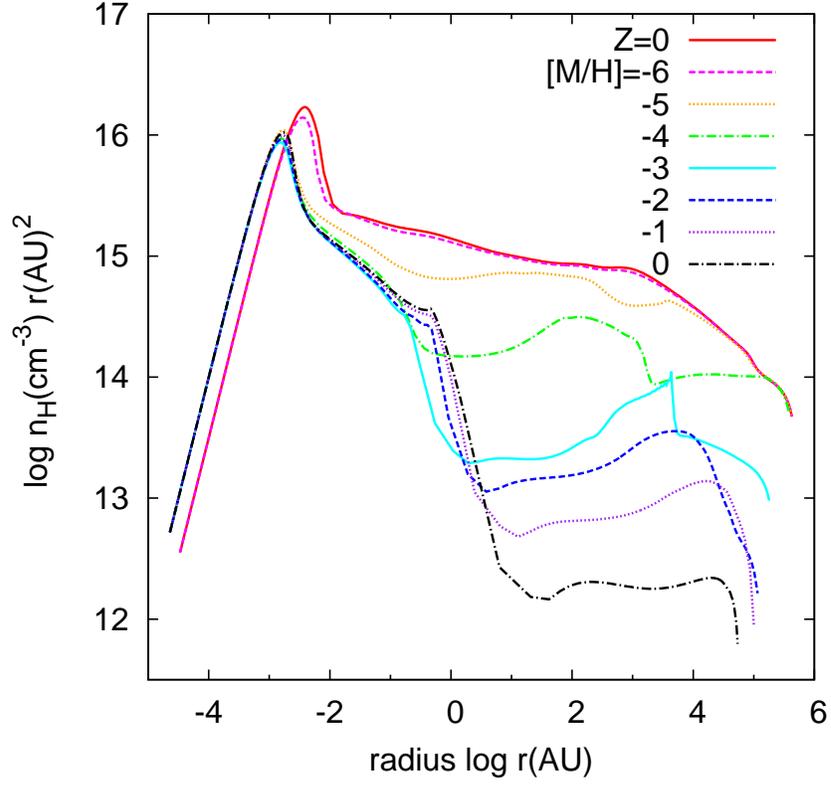} \caption{The radial distributions of
density at the protostar formation epochs for the cloud cores with
different metallicities.  To emphasize differences among the cases,
the number density times radius squared $n_{\rm H}({\rm cm^{-3}})
r^{2}({\rm AU})$ is shown.
\label{fig:rn.all}}
\end{figure}

\begin{figure}
\plotone{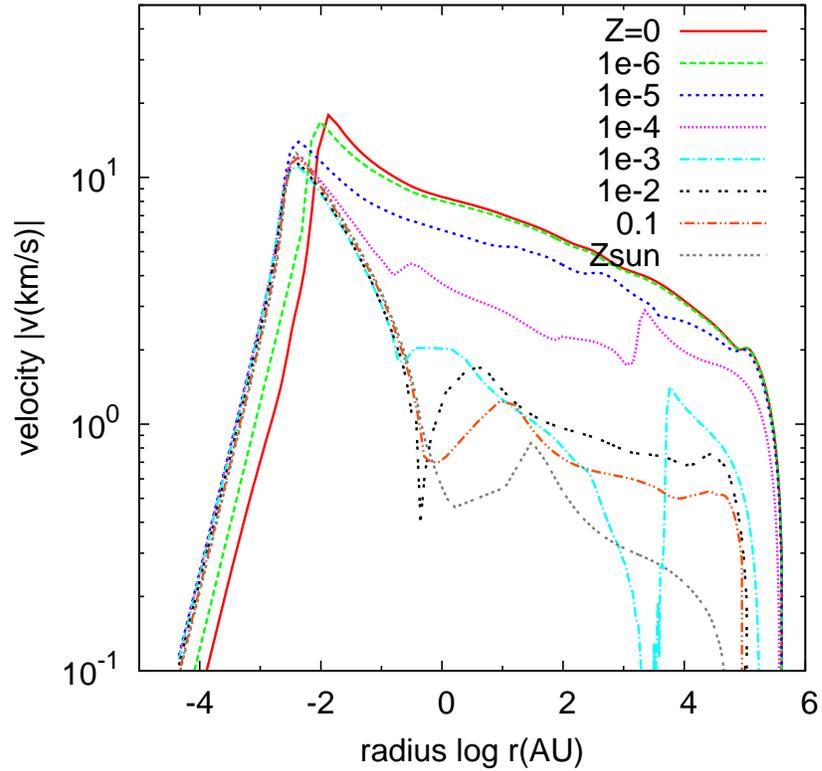} \caption{The radial distributions of
velocity at the protostar-formation epochs for the cloud cores with
different metallicities.
\label{fig:rv.all}}
\end{figure}

\begin{figure}
\plotone{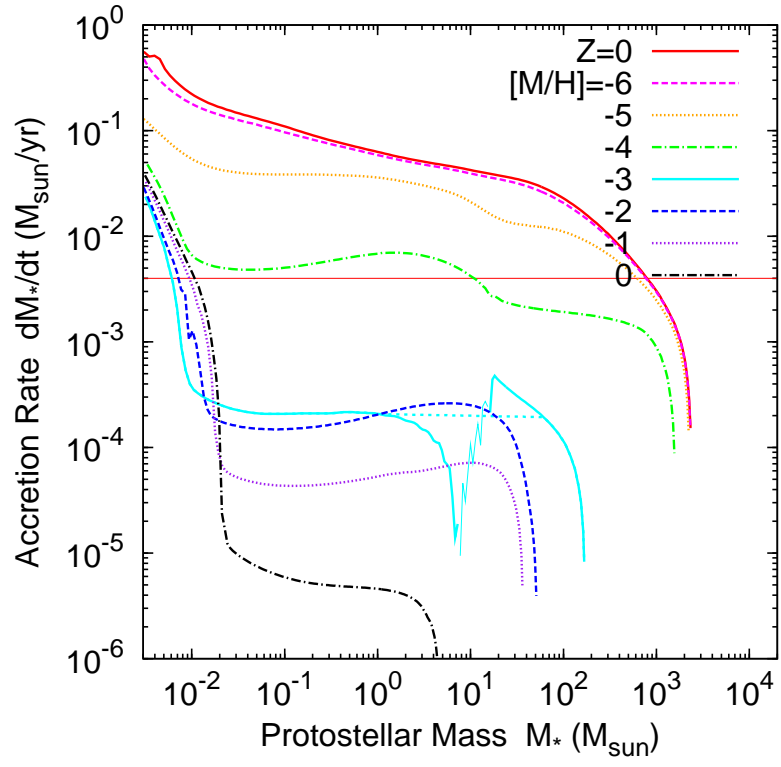} \caption{The mass accretion rates as
functions of the protostellar mass derived by equation
(\ref{eq:mdot}). The horizontal line at $dM_{\ast}/dt=4 \times
10^{-3}M_{\sun}/{\rm yr}$ indicates the value above which the
stationary accretion becomes impossible by the radiation force due to
the accretion luminosity during the Kelvin-Helmholtz contraction.
\label{fig:mdot}}
\end{figure}

When the protostar is not very massive ($\la 10-100M_{\sun}$), stellar
feedback to the accretion flow is not significant yet.  Then the
accretion rate is determined by the density and velocity structures in
the envelope at the moment of protostar formation, which are shown in
Figures \ref{fig:rn.all} and \ref{fig:rv.all}, respectively.  For
density distributions, the number density multiplied by $r^{2}$ is
presented to emphasize differences among the cases with different
metallicities. As a general trend, both the density and velocity are
higher for lower metallicity owing to higher prestellar temperature as
is expected for the LP similarity solution, which has density 4.4
times the singular-isothermal-sphere value $\propto c_{\rm
s}^{2}/r^{2} \propto T/r^{2}$ and infall velocity 3.3 times the sound
speed $c_{\rm s} \propto T^{1/2}$ (Sec. 3.3).  We evaluate the
accretion rate as
\begin{equation}
\dot{M}(M_r)= 4 \pi r^{2} (-v) \rho \bigg|_{M_r}.
\label{eq:mdot}
\end{equation}
These values are shown in Figure \ref{fig:mdot} as a function of the
enclosed mass $M_r$, which can be regarded as the protostellar mass
$M_{\ast}$ at the time when the mass element of interest falls on the
protostar.  It should be noted that the accretion rates shown in
Figure \ref{fig:mdot} are about ten times higher than the often cited
values of $c_{s}^3/G$, which are valid for the accretion from the
static singular isothermal sphere (Shu 1977), because of the dynamical
nature of the collapse.  For the LP solution, this factor is $\simeq
50$. Although the density distribution soon reaches that of LP
solution, convergence of the velocity to the LP solution is slow and
does not reach the value of LP solution during the actual
collapse. This results in a smaller factor of $\sim 10$ enhancement
instead of $\simeq 50$ from $c_{s}^3/G$ (Foster \& Chevalier 1983;
Larson 2003)

In the case of metallicity [M/H]=-3, expansion, rather than infall, is
observed at a small radial range owing to oscillation of the core at
the sudden H$_2$-formation heating. This results in the negative value
of accretion rate by equation (\ref{eq:mdot}; thin line in Figure
\ref{fig:mdot} at $\sim 10 M_{\sun}$) and followed by enhanced rate
later. In this range, the rate does not give the correct accretion
rate.  In the following calculation of protostellar evolution, we use
a rate extrapolated from lower protostellar mass until the accretion
rate becomes the same as this value (dotted line in Figure
\ref{fig:mdot}) in this range.

\begin{figure}
\plotone{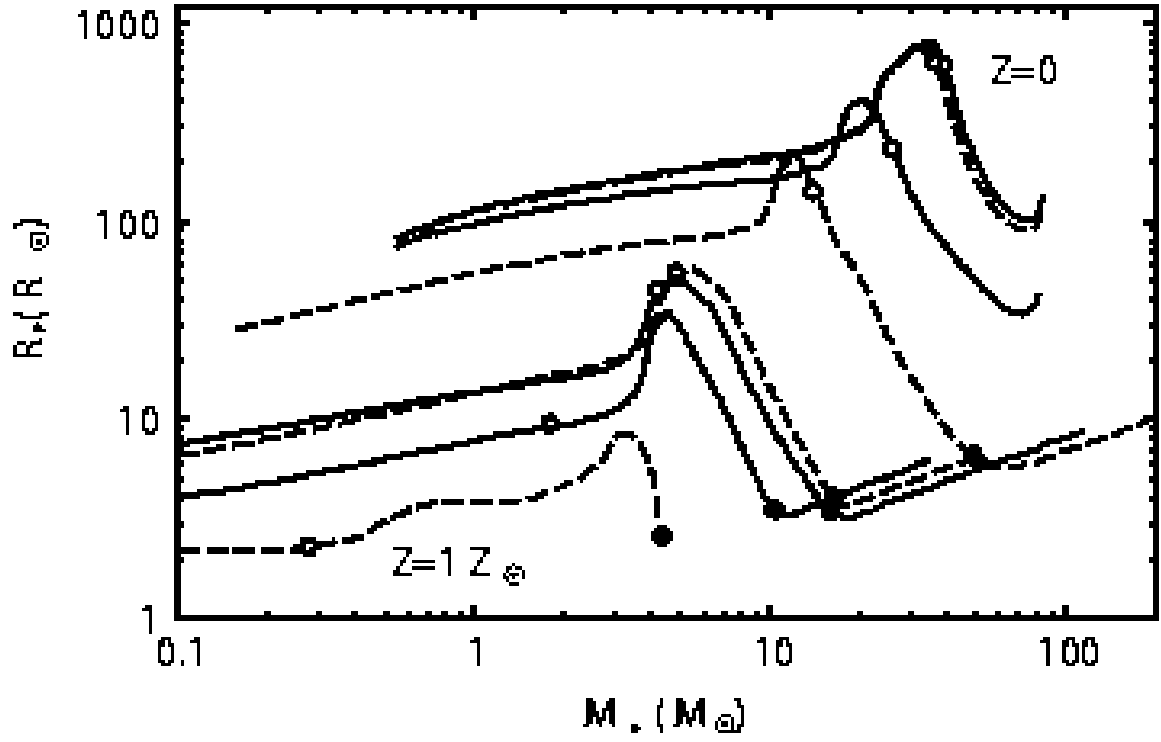} \caption{ The evolution of
protostellar radius during the accretion phase.  The thick (dashed)
lines represent the evolution with different metallicities of $Z =0$,
$10^{-5}~Z_\odot$, $10^{-3}~Z_\odot$, and $10^{-1}~Z_\odot$ in
descending order ($Z=1~Z_\odot$, $10^{-2}~Z_\odot$, $10^{-4}~Z_\odot$,
and $10^{-6}~Z_\odot$ in ascending order). The open circles on the
lines indicate the epoch when the total energy production rate by
deuterium burning reaches 80\% of the steady burning rate $L_{\rm
D,st} \equiv \dot{M} \delta_{\rm D}$, where $\delta_{\rm D}$ is the
energy available from the deuterium burning per unit gas mass.  The
filled circles mark the epoch when the total energy production rate by
hydrogen burning exceeds 80\% of the luminosity at the stellar surface
$L_\ast$.
\label{fig:m_r}}
\end{figure}

With the accretion rate obtained above, we calculate the accretion
evolution of protostars as in Hosokawa \& Omukai (2009a).  In this
method, originally developed by Stahler et al. (1980) and Stahler \&
Palla (1986), a protostar model is constructed by solving ordinary
stellar structure equations assuming the hydrostatic equilibrium,
while the envelope is solved under the assumption of the stationary
accretion with a given accretion rate (see Hosokawa \& Omukai 2009a
for more details).  Figure \ref{fig:m_r} shows the evolution of
protostellar radius with increasing protostellar mass for different
metallicities.  The calculations are initiated with the protostar mass
of $0.01M_{\sun}$.  For [M/H]$\leq$-5, the accretion rate is too high
at the beginning and we cannot find convergent solutions for
protostars by our numerical scheme.  Then we start from a smaller rate
and increases it with protostellar mass until the rate shown in Figure
\ref{fig:mdot} is reached.  For those cases, we show the protostellar
radii only after the correct accretion rates are reached and transient
behaviours disappeared.  Since the accretion rates adopted are
similar, this result resembles closely that in Figure 3 of Hosokawa \&
Omukai (2009b), which shows the cases of $10c_{\rm s}^3/G$, calculated
by temperatures from the one-zone model.  From Figure \ref{fig:m_r},
we first notice that lower metallicity protostars have larger radii
owing to higher accretion rates. Since a short accretion timescale
does not allow accreted matter to cool enough near the stellar surface
before embedded inside, a protostar with higher accretion rate has
higher entropy content and thus larger radius. In all cases in Figure
\ref{fig:m_r}, the protostellar radius initially increases gradually
with addition of newly accreted matter. When the protostar is small in
mass, the opacity due to the free-free absorption is so high inside
the star that the embedded matter has no time to release its heat
(so-called the {\it adiabatic accretion phase}).  The protostar with
solar metallicity ([M/H]=0) has small entropy content and thus the
entropy generation by deuterium burning influences the evolution,
which is observed as an enhanced increase rate of radius at $\simeq
0.4M_{\sun}$.  In other cases, the deuterium burning has no large
impact in their evolution. The gradual increase in radius in the
adiabatic accretion phase is followed by abrupt swelling.  With
increase of protostellar mass, the temperature inside also increases
and the opacity decreases as $\propto T^{-3.5}$.  At some point, the
decreased opacity allows the heat kept in the stellar interior to
propagate outward by way of radiative diffusion. At this moment, the
protostellar radius swells abruptly and thereafter decreases with
throwing away the contained heat (so-called the {\it Kelvin-Helmholtz
(KH) contraction}). Finally, when the heat generated by hydrogen
burning compensates the radiative loss, the star reaches stable
hydrogen burning phase (solid circles in Fig. \ref{fig:m_r}) of the
zero-age main sequence.  If the accretion is not halted by other
mechanism, for example, blow away of the envelope by radiative force
onto the dusty envelope, quenching of mass reservoir in the envelope,
etc, the protostar continues to grow in mass. The radius after the
arrival of zero-age main sequence is larger for the higher metallicity
star since more efficient hydrogen burning by the CN cycle halts the
KH contraction earlier.

For cases with [M/H]$\leq -5$, the total luminosity, i.e. the sum of
the contributions from the accretion $L_{\rm acc}$ and interior one
$L_{\rm int}$, becomes close to the Eddington luminosity during the KH
contraction because of the high accretion rate and decreasing radius;
note that $L_{\rm acc}/L_{\rm Edd} \propto \dot{M}/R_{\ast}$.  At this
moment, the surface layer of the protostar is thrust outward and the
protostellar radius begins to inflate violently. This marks the end of
the stationary accretion. The accretion may stop completely if this
eruption clears away all the matter in the envelope. Another
possibility is that the mass accretion is regulated by this effect. If
the accretion rate is temporarily decreased, the luminosity falls
below the Eddington value.  Thus the accretion may continue in a
sporadic fashion. For other cases, we continue calculation until all
the matter in the parental core is accreted. The arrival to the
main-sequence occurs at higher protostellar mass for lower
metallicity, e. g., at 50$M_{\sun}$ for [M/H]=-4, while at 4$M_{\sun}$
for [M/H]=0.  For [M/H]=0, due to a small amount of the mass reservoir
(4.7$M_{\sun}$), the accretion ceases shortly after the arrival to the
main-sequence.

\section{Discussion}
\subsection{Radiative Transfer Effects }
\begin{figure}
\plotone{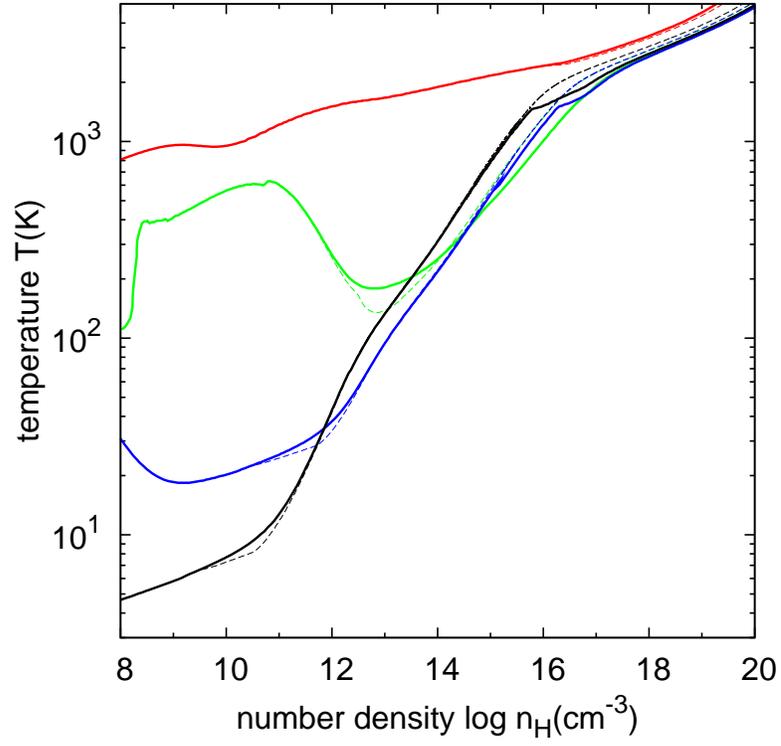} \caption{Radiative transfer effects
on the central temperature evolution. The solid lines are the cases
with radiative transfer (same as Fig.\ref{fig:nTc.all}, but enlarged
for clarity), while dashed lines are the cases with continuum cooling
rate reduced as in equation (\ref{eq:Lesc}).  Although the
temperatures are slightly higher near optically thin to thick
transition regime in the radiative transfer cases, their differences
are small.
\label{fig:nTesc}}
\end{figure}

\begin{figure}
\plotone{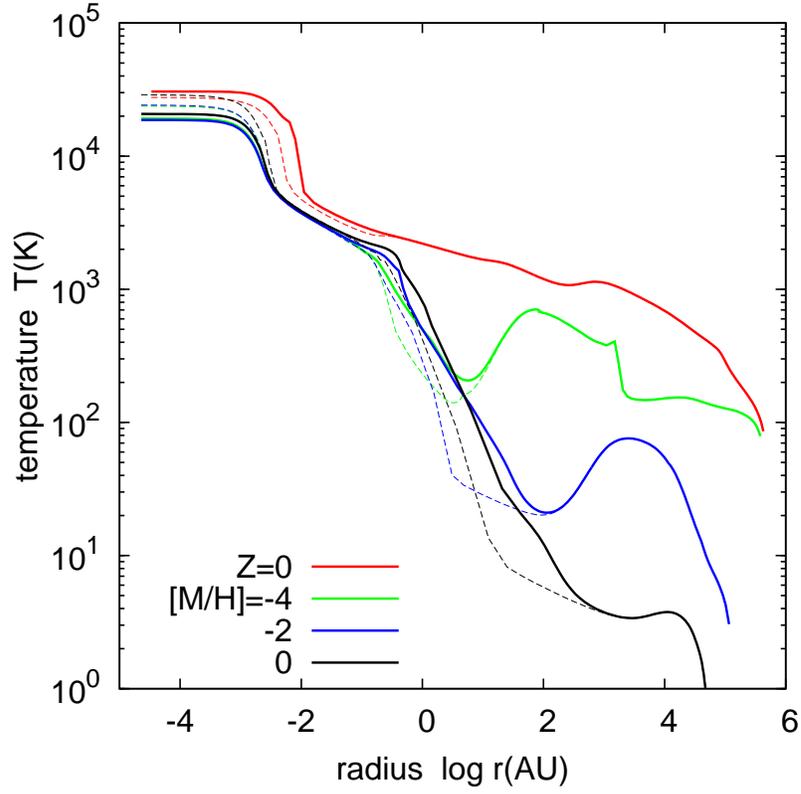} \caption{Radiative transfer effects
on the envelope temperature.  The solid lines are the cases with
radiative transfer, while dashed lines are the cases with continuum
cooling rate reduced as in equation (\ref{eq:Lesc}).  The temperatures
are higher outside the first protostellar cores for the radiative
transfer cases. Note that these differences are more pronouced than
those in the central temperatures (Fig.\ref{fig:nTesc}) owing to
heating from inner hot regions, which is not included in the simple
method by equation (\ref{eq:Lesc}).
\label{fig:rTe.all}}
\end{figure}

In this paper, we have solved radiative transfer for the continuum by
ray tracing method, while in the one-zone model we just reduced the
cooling rate from the optically thin value $\Lambda_{\rm cont, thin}$
by
\begin{equation}
\Lambda_{\rm cont}=\Lambda_{\rm cont, thin} {\rm min}(1, \tau^{-2}),
\label{eq:Lesc}
\end{equation}
where $\tau$ is the optical depth of the continuum absorption.  To see
whether such a simple and local treatment works in hydrodynamical
models, we also calculate the cases where the continuum cooling rate
at radius $r$ is evaluated by using equation (\ref{eq:Lesc}), where
outward radial optical depth $\tau_{\rm r}$ is used for $\tau$. The
evolution of central temperature is shown in Figure \ref{fig:nTesc}
(dashed lines), along with those with radiative transfer (solid lines)
for comparison.  The central evolution is reproduced very well by this
simple treatment although the temperature is slightly lower near the
opacity limit $\tau \simeq 1$ and higher when dust is evaporated at
$\sim 10^{16-17} {\rm cm^{-3}}$.  The former is likely due to our
artificial switching of the cooling rate by equation (\ref{eq:Lesc})
from optically thin to thick cases.  In the radiative transfer cases,
this transition proceeds in a more gradual way since the cooling rate
is somewhat reduced from the thin value for the optical depth $\la 1$.
For the $Z=0$ case, higher cooling rate near the opacity limit at
$\sim 10^{16}{\rm cm^{-3}}$ results in a slightly higher molecular
fraction, which mitigates temperature increase in a later phase and
delays the protostar formation (see Fig. \ref{fig:rTe.all}).  The
higher temperature around the dust evaporation is explained as
follows. In the dust evaporation phase, the central part is already
devoid of dust and transparent, while the surrounding matter is still
optically thick. Then the central part can deposit its heat to the
cooler surroundings by the gas continuum radiation.  In this case, the
rate by equation (\ref{eq:Lesc}) under-estimates the real value since
the radial optical depth $\tau_{\rm r}$, which integrates all the
opacity along the radial direction, is too large. This results in the
higher temperature.

Our ray-tracing method works well to describe the core evolution
accurately.  However, it does not to reproduce radial temperature
distribution very well. The radial temperature distribution at the
protostar formation is presented in Figure \ref{fig:rTe.all} (dashed
lines) for $Z=0$, [M/H]=-4, -2, and 0.  We see clearly deviations in
temperature near the first protostellar core at 1-100AU. In the
radiative transfer model, the temperature are higher owing to
radiative heating by the inner warm gas.  Higher entropy of this
heated gas results in larger size of the first protostellar core in a
later phase than evaluated by the central temperature (e.g., Tomida et
al. 2010).

\subsection{Critical Metallicity for Dust-induced Fragmentation}

\begin{figure}
\plotone{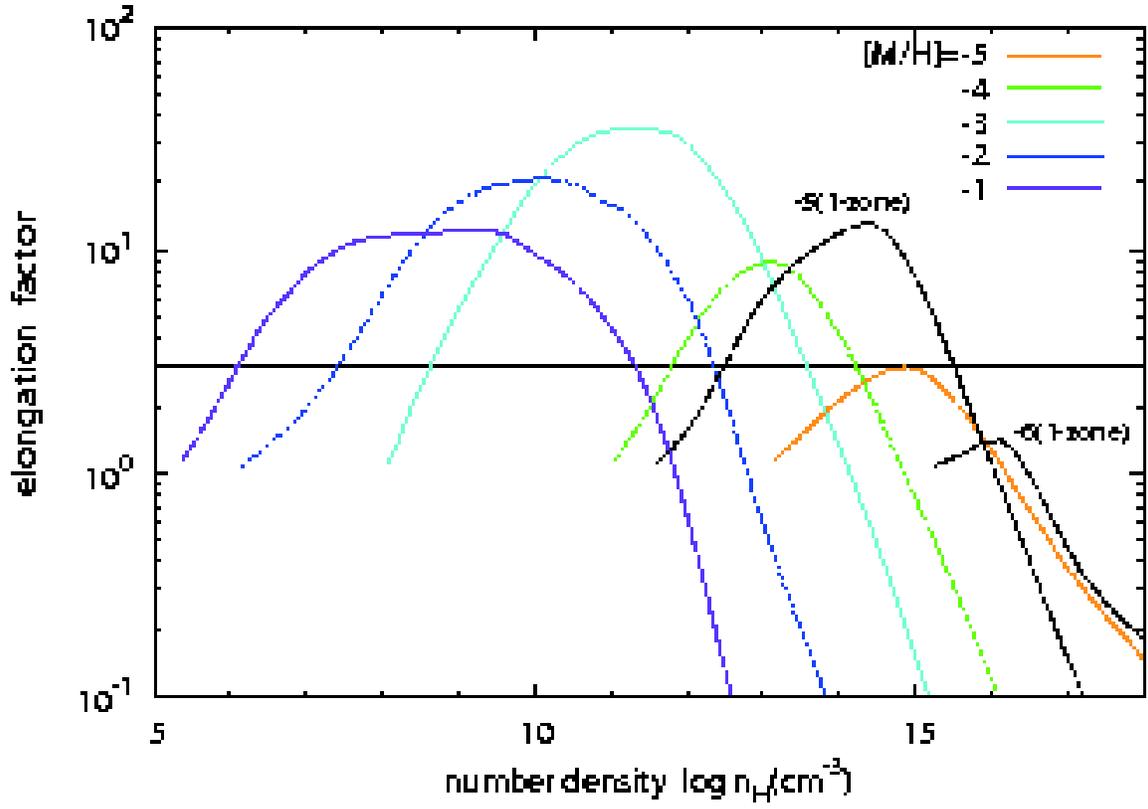} \caption{The growth and decline of
elongation factors during the dust-cooling phase.  The cases of
metallicities [M/H]=-5, -4, -3, -2, and -1 are shown.  For comparison,
the cases [M/H]=-6, and -5 by the one-zone model is also presented.
The dotted horizontal line indicates the critical value $\simeq 3$ for
fragmentation.
\label{fig:elong}}
\end{figure}

Armed with the temperature evolution obtained above, we now discuss
the critical metallicity for fragmentation to produce low-mass
objects.  Tsuribe \& Omukai (2006) studied the fragmentation of dense
cloud cores in the dust-cooling phase by three-dimensional numerical
simulations. They followed the evolution of the core elongation ${\cal
E}=a/b-1$, where $a$ ($b$) is the semi-major (minor) axis.  In their
simulation, a core with ${\cal E} \sim 1$ at the onset of the
dust-cooling phase fragmented to multiple clumps when ${\cal E}$
reached a critical value ${\cal E} \simeq 30$, which corresponds to
$\sim 3$ in the linear theory.  For more details, we refer Tsuribe \&
Omukai (2006). We apply this criterion to our hydrodynamic simulation
as follows.  First, we assume that the cores have modest elongations
of about unity at the beginning of the dust-cooling phase (defined
where $\gamma <1$ and cooling is dominated by the dust). We expect
that the core fragments in this phase when the elongation extrapolated
by the linear theory exceeds 3.  Figure \ref{fig:elong} shows the
elongation factor, which is defined as how many times the elongation
has grown from its initial value for [M/H]=-1...-5. For [M/H]=-6,
there is no clear dust-cooling phase, while for [M/H]=0, rapid
dust-cooling phase with $\gamma <1$ has almost finished by the
beginning of the hydrodynamical calculation at $10^{4}{\rm cm^{-3}}$.
For cases [M/H]$>-5$, the elongation factor well exceeds the critical
value of 3, while for the case with [M/H]=-5, it only marginally
reaches 3.  For lower metallicities, it is clear that the elongation
factor remains below 3.  We thus conclude that for cores initially
elongated with ${\cal E} \sim 1$, the critical metallicity is [M/H]=-5
in our hydrodynamical models. Of course, whether fragmentation
actually occurs depends on our assumption that the elongation is of
order of unity at the beginning of the dust-cooling phase.  In
addition, in the discussion above, we used the temperature evolution
derived by the calculation assuming spherical symmetry. When the core
becomes significantly elongated, the collapse timescale as well as the
cooling timescale will be different, which alters the temperature
evolution.  Note also that the dust-cooling rate depends on the
uncertain dust-gas ratio and dust composition in the early universe
(Schneider et al. 2006).  For example, if the dust-gas ratio is lower
than assumed here, the critical metallicity should be shifted upward
with its inverse proportion. Also more refractory dust species, e.g.,
amorphous carbon rather than organic carbon, can survive in higher
temperature condition of lower metallicity gas, thus enhancing the
dust-cooling rate.  Despite these uncertainties, the value of critical
metallicity derived here provides an approximate estimate for further
studies of low-metallicity star formation.

\subsection{Protostellar Heating to Prevent Fragmentation}

Stellar radiation heats up the surrounding material. For example,
Vazquez-Semadeni et al. (2010) studied numerically the effect of the
stellar ionization feedback from massive stars on the evolution of
giant molecular clouds, and found that it controls their star
formation efficiency.  Since the Jeans mass gives the typical mass of
fragments, this effect suppresses production of many tiny objects as a
result of fragmentation.  In the context of formation of present-day
massive stars, Krumholtz (2006) pointed out that heating by accretion
luminosity from a protostar alleviates fragmentation of the parental
core, which facilitates monolithic collapse to a single massive object
rather than fragmentation into a cluster of small ones.  Krumholtz et
al. (2007) confirmed this effect by raditation hydrodynamical
calculation. Also, Bate (2010) found that radiative feedback from an
accreting protostar suppresses fragmentation of the circumstellar disk
by radiation hydrodynamical simulation. Such a raditive feedback could
be potentially important also in a low-metallicity star forming core,
owing to high accretion rate and thus luminosity of the central
protostar and close separations between the fragments in the
dust-cooling phase.  Here we examine whether this effect suppresses
fragmentation in low-metallicity environments, as studied by Krumholz
(2006) for the present-day star formation.

\begin{figure}
\epsscale{0.8} \plotone{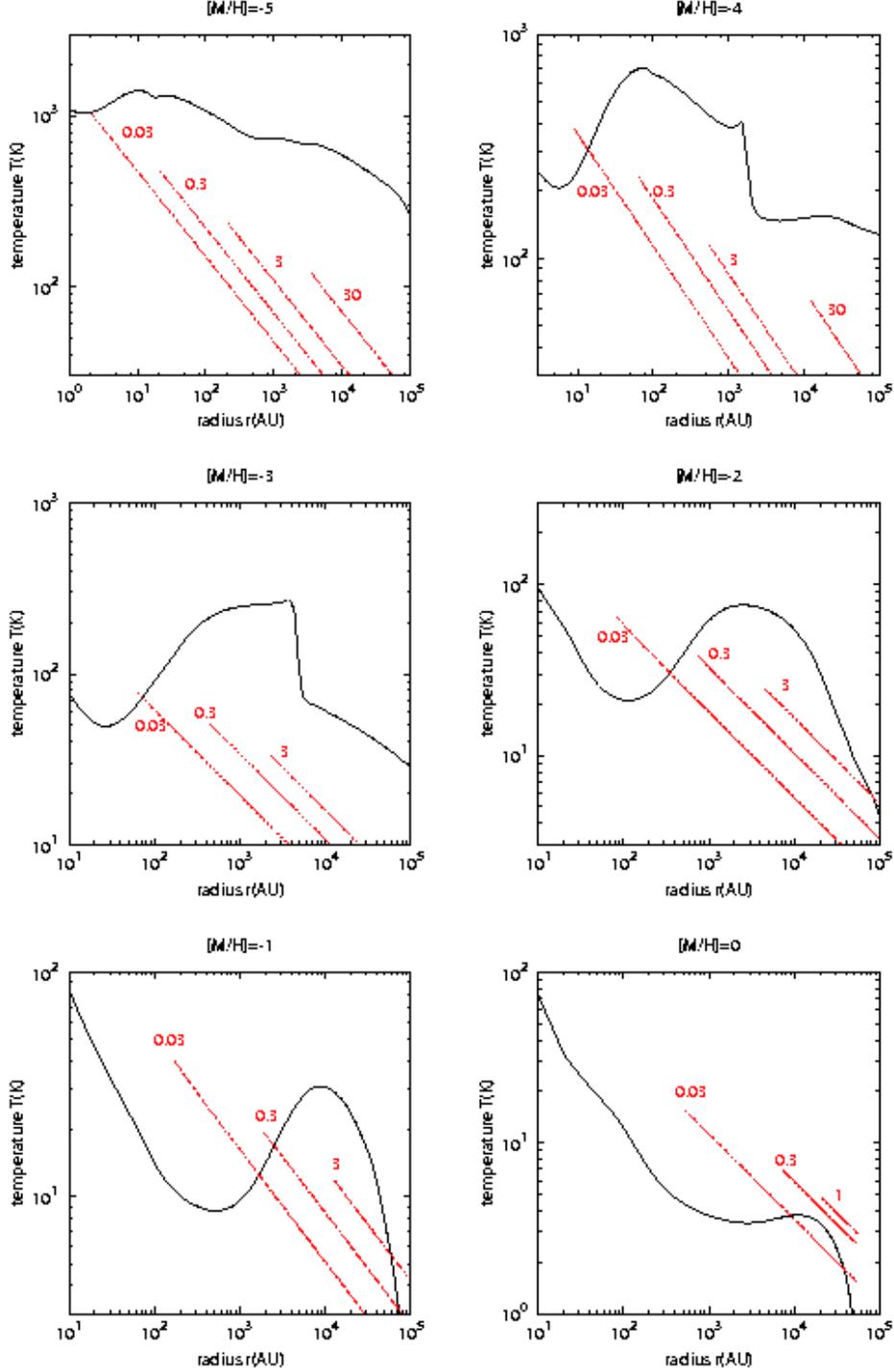} \caption{Heating
effect on the ambient gas by radiation from the central protostar. The
solid lines show the prestellar temperature $T_{\rm pre}$ as a
function of radius. The dashed lines show the dust temperature due to
protostellar radiative heating $T_{\rm heat}=L/16 \pi \sigma r^2$,
where $L$ is the protostellar luminosity. The protostellar masses
$M_{*}$ of those lines are indicated by numbers.  The innermost point
of a dashed line corresponds to the location of the rarefaction wave
at $2M_{*}$, inside of which the matter falls at almost free-fall rate
and the dynamics is not affected by protostellar heating. In cases
with [M/H]$\leq -3$, the lines for $30M_{\sun}$ are not shown because
the total core mass is less than $60M_{\sun}$ and the entire core has
been swept by the rarefaction wave at $M_{*}=30M_{\sun}$. Similarly,
the $T_{\rm heat}$ line at $M_{*}=3M_{\sun}$ is not illustrated in the
$1Z_{\sun}$ case, where that at $M_{*}=1M_{\sun}$ is shown instead.
In the region where $T_{\rm pre}$ is higher than $T_{\rm heat}$, the
protostellar heating effect is negligible.  Since the actual dust
temperature $T_{\rm dust}$ is determined by the balance between the
cooling by its thermal radiation and heating not only by the
protostellar radiation but also by the collisional coupling with gas,
the dust temperature in the accretion phase roughly coincides with the
higher of $T_{\rm heat}$ or $T_{\rm pre}$. If $T_{\rm dust} > T_{\rm
pre}$ (i.e., $T_{\rm heat}>T_{\rm pre}$), the gas temperature
approaches $T_{\rm dust}$ ($\simeq T_{\rm heat}$ in this case) by the
collisional coupling.
\label{fig:rT.fb}}

\end{figure}

In the envelope of a protostar radiating at luminosity $L$, the dust
is heated to a temperature $T_{\rm heat} = \left(L/16 \pi \sigma r^{2}
\right)^{1/4}$, which comes from the condition of radiative
equilibrium.  Note that this is just a measure for heating effect and
is neither the actual dust nor gas temperatures: the true values of
those temperatures are determined by taking account of the collisional
coupling between the gas and dust.  In the following, we compare the
gas temperature $T_{\rm env}$ in the envelope before the heating and
$T_{\rm heat}$, which is the maximum temperature by the heating effect
in the region where $T_{\rm heat}>T_{\rm env}$.  This dust temperature
$T_{\rm heat}$ is shown in Figure \ref{fig:rT.fb} for ${\rm log}
(M_{\ast}/M_{\sun})=-1.5, -0.5, 0.5$ and 1.5.  The innermost radii of
the lines for $T_{\rm heat}$ correspond to the positions of
rarefaction waves. During the accretion phase, the inner matter falls
onto the protostar earlier. The rarefaction front propagates outward
with a sound speed and locates at $\sim 2M_{\ast}$ in terms of the
Lagrangian mass coordinate at the time of protostellar mass $M_{\ast}$
for the density distribution $\rho \propto r^{-2}$.  The matter inside
of the rarefaction front is falling at about a free-fall speed and
probably is not affected by the protostellar heating anymore.  In
Figure \ref{fig:rT.fb}, we see that $T_{\rm heat}$ at a fixed radius
increases with the protostellar mass due to increasing luminosity.
However, the propagation of the rarefaction front is faster than
increase of heating effect: the temperature at the rarefaction front
decreases with increasing protostellar mass. Also, the effect of
heating is less pronounced for lower-metallicity protostars.  Below
[M/H] $\leq -3$, $T_{\rm heat}$ falls below $T_{\rm env}$ at anytime
outside the rarefaction front. Namely, the protostellar heating has no
impact on the evolution because of the already high prestellar
temperature. For [M/H]=-2 and -1, the radiative heating may fill the
temperature dip due to the dust cooling, but this phase lasts only for
a short period. By $M_{\ast}=0.3M_{\sun}$, this radiative heating
effect becomes negligibly small.  Only for [M/H]=0, the temperature
due to the radiative heating is higher than the prestellar value
throughout the evolution.

The tendency that radiative heating has more impact for higher
metallicity (i.e., lower prestellar temperature) and for lower
protostellar mass can be understood in the following way. When a
protostar is not massive and still in the adiabatic phase, i.e.,
$M_{\ast} \la 10M_{\sun}$, its radius is approximated as (Stahler \&
Palla 1986)
\begin{equation}
R_{\ast}=48.1R_{\sun} \left( \frac{M_{\ast}}{M_{\sun}} \right)^{0.27}
\left( \frac{\dot{M}}{4.41\times10^{-3}M_{\sun}/{\rm yr}}
\right)^{0.41} .
\end{equation}
We write the accretion rate as
\begin{eqnarray}
\dot{M} &=& \phi \frac{c_{\rm s}^{3}}{G} \\ &=& 1.63 \times 10^{-5}
M_{\sun}/{\rm yr}~\phi_{10}~T_{10}^{3/2},
\end{eqnarray}
by using a numerical factor $\phi$ ($\sim 10$ in our case), and
$\phi_{10}=\phi/10$ and $T_{10}=T/10{\rm K}$.  In the adiabatic phase,
the protostellar luminosity is dominated by the contribution from the
accretion luminosity:
\begin{eqnarray}
L \simeq L_{\rm acc} & \simeq & \frac{G M_{\ast} \dot{M}}{R_{\ast}} \\
& = & 1.01 \times 10^{2} L_{\sun}~\phi_{10}^{0.59}~T_{10}^{0.89} ~
m_{\ast}^{0.73},
\end{eqnarray}
where $m_{\ast}=M_{\ast}/M_{\sun}$.  Thus, temperature of dust heated
by the protostar is
\begin{eqnarray}
T_{\rm heat} &=& \left( \frac{L_{\rm acc}}{16 \pi \sigma r^{2}}
\right)^{1/4} \\ &=& 2.82 \times 10^{2} {\rm
K}~\phi_{10}^{0.15}~T_{10}^{0.22} ~m_{\ast}^{0.18}~r_{10}^{-1/2},
\end{eqnarray}
where $r_{10}=r/10{\rm AU}$.  Suppose the location of rarefaction
front $M_{\rm rw}=2 M_{\ast}$ and density distribution is a factor of
$\alpha$ times that of the singular isothermal sphere $\rho=\alpha
\rho_{\rm SIS}$.  For the Larson-Penston collapse $\alpha \simeq 4.4$,
and we set it a representative value: $\alpha_{4.4}=\alpha/4.4$.
Using the radius of the rarefaction front
\begin{eqnarray}
r_{\rm rw} &=& \frac{G M_{\ast}}{\alpha c_{\rm s}^{2}} \\ &=& 5.59
\times 10^{3} {\rm AU}~\alpha_{4.4}^{-1}~T_{10}^{-1}~m_{\ast}.
\end{eqnarray}
We obtain the dust temperature due to protostellar heating at the
rarefaction front
\begin{equation}
T_{\rm heat}(r_{\rm rw})= 11.9 {\rm
K}~\alpha_{4.4}^{0.5}~\phi_{10}^{0.15} ~
T_{10}^{0.72}~m_{\ast}^{-0.32}.
\end{equation}
Therefore, the condition on the protostellar mass for $T_{\rm
heat}(r_{\rm rw})$ to exceed the prestellar temperature is given by
\begin{equation}
M_{\ast} < 1.7
M_{\sun}~\alpha_{4.4}^{1.57}~\phi_{10}^{0.46}~T_{10}^{-0.88}.
\label{eq:mheat}
\end{equation}
This demonstrates that the radiative heating effect is important only
up to some mass given by equation (\ref{eq:mheat}), which decreases
rapidly with increasing prestellar temperature, as has been discussed
from Figure \ref{fig:rT.fb}.

In the course of the accretion, circumstellar disks form owing to the
angular momentum. The gravitational instability of rapidly accreting
disks were investigated by Kratter et al. (2010) for the isothermal
gas by way of global numerical experiments. They found that the
central parameter for the stability of an accreting disk is $\xi
\equiv \dot{M}/(c_{\rm s, d}^3/G)=\phi (c_{\rm s}/c_{\rm s, d})^3$,
which parameterized the accretion rate. Here $c_{\rm s, d}$ is the
sound speed in the disk and $c_{\rm s}$ is that in the parental core,
i.e., before the matter falls on the disk.  Rapidly accreting disks
with $\xi \ga 3$ become unstable and fragment into multiples or
binaries, although the exact value for threshold depends on a
rotational parameter of the parental cores, which stabilizes the disks
somewhat.  In our cases, since $\phi \sim 10$, the disks will become
unstable and fragment without significant heating.  Being closer to
the star, cicumstellar disks can be more vulnerable to the radiative
feedback than the matter outside the rarefaction wave, which has been
discussed above. Although not discussed here, the radiative feedback
onto the disk is interesting in this respect and awaits further
investigations in future.

\section{Summary}
We have studied star formation with different metallicities,
$0-1Z_{\sun}$, under the assumption of spherical symmetry.  Until the
birth of hydrostatic protostars, the gravitational collapse of a dense
cloud core is calculated using chemo-radiation hydrodynamics.  The
subsequent protostellar accretion is followed by assuming the
accretion flow is stationary while ordinary stellar structure
equations are solved for the central protostar.

Our findings are summarized as follows:

(1) Prestellar temperature is higher in a lower-metallicity cloud core
during optically thin stage, while it approaches a single
density-temperature relation after the core becomes optically
thick. Only in lowest metallicity ($\leq 10^{-6}Z_{\sun}$) cases, the
temperature remains slightly higher than other cases.

(2) Since the temperature evolution converges in cases with
metallicity $\geq 10^{-5}Z_{\sun}$, physical conditions of protostars
at their formation are universal with a mass of $10^{-3} M_{\sun}$, a
radius of $0.5R_{\sun}$.  The density and temperatures are $\sim
10^{22}{\rm cm^{-3}}$ and $3\times 10^{4}$K respectively.  In the
cases with $\leq 10^{-6}Z_{\sun}$, owing to the slightly higher
prestellar temperature, the protostar formation occurs about an order
of magnitude earlier in density at $\sim 10^{21}{\rm cm^{-3}}$, with
slightly lower temperature $2\times 10^{4}$K, and somewhat higher mass
$2\times 10^{-3} M_{\sun}$ and radius $1R_{\sun}$.
 
(3) The prestellar collapse approaches the Larson-Penston self-similar
solution, for which the velocity is 3.3 times the sound speed $c_{\rm
s}$ and the density is 4.4 times the singular isothermal sphere value
($\propto c_{\rm s}^{2}$).  Reflecting the higher prestellar
temperature, the velocity and the density in the envelope of a
lower-metallicity dense core is higher at a given radius.

(4) The temperature evolution at the cloud center is roughly
consistent with that found in our previous one-zone models, although
collapse is faster than assumed in the one-zone model, which results
in a slightly higher temperature at the center. The faster collapse,
in particular, affects the temperature in the dust-cooling
phase. Since the fragmentation into low-mass objects occurs in this
phase, the critical value of metallicity for inducing the
fragmentation can be altered by this effect.

(5) The accretion rate after the birth of the protostar is higher in
the lower-metallicity case because of the relation between accretion
rate $\dot{M}$ and the prestellar sound speed $c_{\rm s}$ in
similarity solutions, $\dot{M} = \phi c_{\rm s}^{3}/G$, where $\phi$
is a non-dimensional constant.  For the isothermal Larson-Penston
solution, which describes a highly dynamical collapse, $\phi=46.9$,
while for the Shu solution, which is valid for accretion from a static
isothermal sphere, $\phi=0.975$. In our cases, owing to modestly
dynamical nature of the collapse, the accretion rates fall in between
these two extremes and $\sim 10c_{\rm s}^{3}/G$.

(6) The evolution of protostars under the accretion rates obtained
from the prestellar collapse calculations is followed under the
assumption of stationary accretion flows. The protostellar mass at its
main sequence arrival is higher in the lower-metallicity case owing to
the higher accretion rate: for example, it is $4M_{\sun}$ for
$1Z_{\sun}$, while $50M_{\sun}$ for $10^{-4}Z_{\sun}$.  For $\leq
10^{-5}Z_{\sun}$, too high an accretion rate causes the luminosity to
reach the Eddington limit and the accretion is almost halted at
$80M_{\sun}$. No stationary solution exists thereafter. The accretion
may be halted temporarily. Once the accretion rate is reduced,
however, the luminosity falls below the Eddington limit and thus the
accretion would be resumed. The accretion may continue in such a
non-stationary or a sporadic fashion.

(7) The critical metallicity for low-mass fragmentation is evaluated
by applying the criterion by Tsuribe \& Omukai (2004) to the
calculated temperature evolution at the center.  With metallicity $Z
\ga 10^{-5}Z_{\sun}$, a moderately elongated cloud core is likely to
fragment into low-mass objects during the dust-cooling phase.

(8) Radiation from a forming protostar heats up its ambient medium,
thereby possibly preventing the fragmentation in the surrounding
region.  This protostellar heating effect is evaluated for different
metallicity cases.  For low-metallicities $\la 10^{-2}Z_{\sun}$, the
prestellar temperature is already high, and thus the heating effect is
turned out to be insignificant.  For higher metallicities,
protostellar heating can be important especially in the early
protostellar phase.

Effects such as rotation, magnetic fields, and turbulence, which are prevalent in the
nearby star-forming clouds, have not been addressed in our spherical symmetric analysis.
They would delay the contraction during the collapse phase and 
modify the accretion rate in the subsequent phase. 
In addition, rotation facilitates the fragmentation by forming a thin disk, 
while magnetic fields counteract it by extracting the angular momentum (Machida et
al. 2008; Hennebelle \& Teyssier 2008).
Those effects surely await future studies. 

Overall, our calculations clarified the similarlity and differences in
the formation of protostars with different metallicities. While some
of the important issues, such as the critical metallicity for
fragmentation, need further studies using full three-dimensional
simulations, our results provide a foundation for such subsequent
studies. Low-metallicity star formation is particularly important in
the context of the formation of premeval galaxies. Most of the planned
ground-based and space-borne telescopes are aimed at observing the
first galaxies in which the gas metallicity is supposed to be very
low. Theoretical studies such as those presented here will give an
invaluable input for the future observations.

{\acknowledgements The present work is supported in part by the
Grants-in-Aid by the Ministry of Education, Science and Culture of
Japan (19047004, 2168407, 21244021:KO, 20674003:NY). 
}

\bibliographystyle{apj}
\bibliography{refs}

\clearpage

\appendix

\section{Reduced chemical network}
The set of reactions are presented in Table 1. The number of reactions
and rate coefficients are the same as Table 1 of Omukai et al. (2005).
This model includes 24 compounds of hydrogen, deuterium, carbon and
oxygen, i.e., ${\rm H^+}$, ${\rm e}$, ${\rm H}$, ${\rm H_2}$, ${\rm
D^+}$, ${\rm D}$, ${\rm HD}$, ${\rm C^+}$, ${\rm C}$, ${\rm CO}$,
${\rm CO_2}$, ${\rm O}$, ${\rm OH}$, ${\rm H_2O}$, ${\rm O_2}$ ${\rm
H^-}$. Non-equilibrium reactions are solved for the following 15 major
species: ${\rm H^+}$, ${\rm e}$, ${\rm H}$, ${\rm H_2}$, ${\rm D^+}$,
${\rm D}$, ${\rm HD}$, ${\rm C^+}$, ${\rm C}$, ${\rm CO}$, ${\rm
CO_2}$, ${\rm O}$, ${\rm OH}$, ${\rm H_2O}$, ${\rm O_2}$ (see Table
1).  The other 9 species (${\rm H^-}$, ${\rm CH}$, ${\rm CH_2}$, ${\rm
CO^+}$, ${\rm O^+}$, ${\rm OH^+}$, ${\rm H_2O^+}$, ${\rm H_3O^+}$,
${\rm O_2^+}$), which appears only as intermediaries of the reactions
among the above major species, are set to be chemical equilibrium as
follows:
\begin{equation}
{\rm H^-}: k_{2}n(H)n(e)=k_{3}n(H^-)n(H)
\label{eq:A1}
\end{equation}
\begin{equation}
{\rm CH}: [k_{37}n(H)+k_{28}n(H_2)]n(C)+k_{22}n(CH_2)n(H)
=n(CH)[k_{21}n(H)+k_{31}n(H_2)+k_{42}n(O)]
\end{equation}
\begin{equation}
{\rm CH_2}: k_{62}n(H_2)n(C)+k_{31}n(H_2)n(CH)=k_{22}n(CH_2)n(H)
\end{equation}
\begin{equation}
{\rm CO^+}: k_{48}n(C^+)n(OH)=k_{54}n(CO^+)n(H)
\end{equation}
\begin{equation}
{\rm O^+}:
k_{30}n(H^+)n(O)+k_{49}n(C^+)n(O_2)=[k_{50}n(H)+k_{51}n(H_2)]n(O^+)
\end{equation}
\begin{equation}
{\rm OH^+}:
k_{51}n(O^+)n(H_2)+k_{45}n(H^+)n(OH)=[k_{52}n(H_2)+k_{56}n(e)]n(OH^+)
\end{equation}
\begin{equation}
{\rm H_2O^+}: k_{52}n(OH^+)n(H_2)+k_{46}n(H^+)n(H_2O)
=[k_{53}n(H_2)+(k_{57}+k_{58})n(e)]n(H_2O^+)
\end{equation}
\begin{equation}
{\rm H_3O^+}: k_{53}n(H_2O^+)n(H_2)=(k_{59}+k_{60})n(e)n(H_3O^+)
\end{equation}
\begin{equation}
{\rm O_2^+}: k_{47}n(H^+)n(O_2)=k_{61}n(e)n(O_2^+).
\label{eq:A9}
\end{equation}
From (\ref{eq:A1})-(\ref{eq:A9}), the chemical equilibrium values are
\begin{equation}
y(H^-)= \frac{k_{2}y(e)}{k_{3}}
\end{equation}
\begin{eqnarray}
y(CH)&=&\frac{k_{37}y(C)y(H)+k_{28}y(C)y(H_2)+k_{62}y(C)y(H_2)}
	{k_{21}y(H)+k_{42}y(O)} \\
	y(CH_2)&=&\frac{k_{62}y(C)y(H_2)+k_{31}y(H_2)y(CH)}
	{k_{22}y(H)}
\end{eqnarray}
\begin{equation}
y(CO^+)=\frac{k_{48}y(C^+)y(OH)}{k_{54}y(H)}
\end{equation}
\begin{equation}
y(O^+)=\frac{k_{30}y(H^+)y(O)+k_{49}y(C^+)y(O_2)}{k_{50}y(H)+k_{51}y(H_2)}
\end{equation}
\begin{equation}
y(OH^+)=\frac{k_{51}y(O^+)y(H_2)+k_{45}y(H^+)y(OH)}
		{k_{52}y(H_2)+k_{56}y(e)}
\end{equation}
\begin{equation}
y(H_2O^+)=\frac{k_{52}y(OH^+)y(H_2)+k_{46}y(H^+)y(H_2O)}
			{k_{53}y(H_2)+(k_{58}+k_{57})y(e)}
\end{equation}
\begin{equation}
y(H_3O^+)=\frac{k_{53}y(H_2O^+)y(H_2)}{(k_{60}+k_{59})y(e)}
\end{equation}
\begin{equation}
y(O_2^+)=\frac{k_{47}y(H^+)y(O_2)}{k_{61}y(e)} .
\end{equation}

By using the one-zone models, we have compared the results by the
chemical model above and those by fully non-equilibrium calculations
with the same set of reactions and have found that the evolution of
temperature, as well as abundances of the coolants abundances agree
excellently.

\section{CO and OH cooling functions}
We compute the cooling rate of CO and OH rotational transitions by
counting the level populations under the assumption of statistical
equilibrium (see Omukai 2001 Appendix B).  The level energies and
transitional coefficients are provided by Leiden Atomic and Molecular
Database (Schoier et al. 2005).  The dependence of the results for
molecule M on the three variables, the H$_2$ number density $n({\rm
H_2})$, column density parameter $\tilde{N}({\rm M})$ and temperature
$T$ is fitted by the following four-parameter expression according to
Neufeld \& Kaufman (1993):
\begin{equation}
\frac{1}{L_{\rm M}}=\frac{1}{L_{0}}+\frac{n({\rm H_2})}{\cal L_{\rm
LTE}} +\frac{1}{L_{0}} \left[ \frac{n({\rm H_2})}{n_{1/2}}
\right]^{\alpha} \left( 1-\frac{n_{1/2}L_{0}}{\cal L_{\rm LTE}}
\right).
\end{equation}
The column density parameter is defined as $\tilde{N}(M)
=N(M)/\sqrt{\frac{2kT}{m({\rm M})}}({\rm cm^{-2}/km~s^{-1}})$, where
$N({\rm M})$ is the column density, and $m({\rm M})$ is the mass of
molecule M. The fitting parameters $L_{0}$, which is a function of $T$
alone, and ${\cal L_{\rm LTE}}$, $n_{1/2}$, and $\alpha$, which are
functions of both $\tilde{N}(M)$ and $T$, are presented in Table
\ref{table:CO} for CO and in Table \ref{table:OH} for OH.

\clearpage

\newpage
\bigskip
\begin{deluxetable}{cll}
\tablecaption{Reduced Chemical Reactions}
\tablehead{
\colhead{Number} &
\colhead{Reaction} & 
} 
\startdata
1(H1)& ${\rm H^{+}    +   e     \rightarrow   H     +   \gamma}$ \nl
2(H2)& ${\rm H     +   e     \rightarrow   H^-    +   \gamma}$ \nl
3(H3)& ${\rm H^-    +   H     \rightarrow   H_2    +   e}$ \nl
4(H4)&  ${\rm H_2    +   H     \rightarrow 3 H}$ \nl
5(H5)&  $ {\rm 3 H \rightarrow   H_2    +   H}$  \nl
6(H6)&  ${\rm 2 H     +   H_2    \rightarrow 2 H_2}$ \nl
7(H7)&  ${\rm 2 H_2   \rightarrow 2 H     +   H_2}$ \nl
8(H8)&  ${\rm 2 H     +   grain \rightarrow   H_2}$ \nl
11(D1)&  ${\rm D     +   H^+    \rightarrow   D^+     +   H }$ \nl
12(D2)&  ${\rm D^+    +   H     \rightarrow   D     +    H^+ }$ \nl
13(D3)&  ${\rm D     +   H_2    \rightarrow  D     +   HD }$ \nl
14(D4)&  ${\rm D^+    +   H_2   \rightarrow  H^+  +    HD }$ \nl
15(D5)&  ${\rm HD     +   H    \rightarrow   H_2     +   D }$ \nl
16(D6)&  ${\rm HD    +   H^+     \rightarrow H_2     +   D^+ }$ \nl
21(M1)&  ${\rm H     +   CH    \rightarrow   C     +   H_2 }$\nl
22(M2)&   ${\rm H     +   CH_2   \rightarrow   CH    +   H_2 }$\nl
25(M5)&  ${\rm H     +   OH    \rightarrow   H_2    +   O}$ \nl
26(M6)&  ${\rm   H     +   H_2O   \rightarrow   OH    +   H_2}$ \nl
27(M7)&  ${\rm   H     +   O_2    \rightarrow    OH    +   O}$ \nl
28(M8)&  ${\rm   C     +   H_2    \rightarrow    CH    +   H}$\nl
29(M9)&  ${\rm   O     +   H_2    \rightarrow    OH    +   H}$ \nl
30(M10)&  ${\rm   H^+    +   O     \rightarrow    O^+    +   H}$ \nl
31(M11)& ${\rm   H_2    +   CH    \rightarrow    CH_2   +   H}$\nl
34(M14)&  ${\rm   H_2    +   OH    \rightarrow    H_2O   +   H}$ \nl
35(M15)& ${\rm 2 OH              \rightarrow   H_2O   +   O}$\nl
36(M16)& ${\rm   OH    +   CO    \rightarrow   CO_2   +   H}$\nl
37(M17)& ${\rm   C     +   H     \rightarrow   CH    +   \gamma}$\nl
38(M18)& ${\rm   C     +   OH    \rightarrow   CO    +   H}$\nl
39(M19)& ${\rm   C     +   O_2    \rightarrow   CO    +   O  }$  \nl
40(M20)& ${\rm   O     +   H     \rightarrow   OH    +   \gamma}$ \nl
41(M21)& ${\rm 2 O               \rightarrow   O_2    +   \gamma}$ \nl
42(M22)& ${\rm   O     +   CH    \rightarrow   CO    +   H}$\nl
44(M24)& ${\rm   O     +   OH    \rightarrow   O_2    +   H}$ \nl
45(M25)& ${\rm   H^+    +   OH    \rightarrow   OH^+   +   H}$ \nl
46(M26)& ${\rm   H^+    +   H_2O   \rightarrow   H_2O^+  +   H}$ \nl
47(M27)& ${\rm   H^+    +   O_2    \rightarrow   O_2^+   +   H}$ \nl
48(M28)& ${\rm   C^+    +   OH    \rightarrow   CO^+   +   H}$\nl
49(M29)& ${\rm   C^+    +   O_2    \rightarrow   O^+    +   CO}$\nl
50(M30)& ${\rm   O^+    +   H     \rightarrow   H^+    +   O}$ \nl
51(M31)& ${\rm   O^+    +   H_2    \rightarrow   OH^+   +   H}$ \nl
52(M32)& ${\rm   OH^+   +   H_2    \rightarrow   H_2O^+  +   H}$ \nl
53(M33)& ${\rm   H_2O^+  +   H_2    \rightarrow   H_3O^+  +   H}$ \nl
54(M34)& ${\rm   CO^+   +   H     \rightarrow   H^+    +   CO}$\nl
55(M35)& ${\rm   C^+    +   e     \rightarrow   C     +   \gamma}$\nl
56(M36)& ${\rm   OH^+   +   e    \rightarrow    O  +   H}$ \nl
57(M37)& ${\rm   H_2O^+  +   e     \rightarrow   OH    +   H}$ \nl
58(M38)& ${\rm   H_2O^+  +   e     \rightarrow   O     +   H_2}$ \nl
59(M39)& ${\rm   H_3O^+  +   e     \rightarrow   H_2O   +   H}$ \nl
60(M40)& ${\rm   H_3O^+  +   e     \rightarrow   OH    + 2 H}$ \nl
61(M41)& ${\rm   O_2^+   +   e     \rightarrow 2 O}$\nl
62(M42)&  ${\rm   H_2 +   C     \rightarrow   CH_2    +   \gamma}$\nl
\enddata
\end{deluxetable}

\begin{table}
\begin{center}
\renewcommand{\arraystretch}{1.0}
\begin{tabular}{cccccccccccc} \hline
${\rm log_{10}}(\tilde{N})$ &&&& tem&pera&ture \\
\hline
 & 10K & 20K & 30K & 50K & 80K & 100K & 300K & 600K & 1000K & 1500K & 2000K \\
\hline
&&&$-$&${\rm log}_{10}$&$(L_{0}$ &$/{\rm ergs}$& ${\rm cm^3}$&${\rm s^{-1}})$\\
\hline  
&24.77& 24.38& 24.21 & 24.03 & 23.89 & 23.82 & 23.42 & 23.13 & 22.91 & 22.63 
& 22.28 \\
\hline
&&& $-$&${\rm log}_{10}$&$({\cal L_{\rm LTE}}$&$/{\rm ergs}$& ${\rm cm^3}$&${\rm s^{-1}})$\\
\hline 
14.0& 21.08& 20.35& 19.94& 19.45& 19.01& 18.80& 17.81& 17.23& 16.86& 16.66& 16.55 \\
14.5& 21.09& 20.35& 19.95& 19.45& 19.01& 18.80& 17.81& 17.23& 16.86& 16.66& 16.55 \\
15.0& 21.11& 20.37& 19.96& 19.46& 19.01& 18.80& 17.81& 17.23& 16.86& 16.66& 16.55 \\
15.5& 21.18& 20.40& 19.98& 19.47& 19.02& 18.81& 17.82& 17.23& 16.87& 16.66& 16.55 \\
16.0& 21.37& 20.51& 20.05& 19.52& 19.05& 18.83& 17.82& 17.23& 16.87& 16.66& 16.55 \\
16.5& 21.67& 20.73& 20.23& 19.64& 19.13& 18.90& 17.85& 17.25& 16.88& 16.67& 16.56 \\
17.0& 22.04& 21.05& 20.52& 19.87& 19.32& 19.06& 17.92& 17.28& 16.90& 16.69& 16.58 \\
17.5& 22.44& 21.42& 20.86& 20.19& 19.60& 19.33& 18.08& 17.38& 16.97& 16.75& 16.63 \\
18.0& 22.87& 21.82& 21.24& 20.55& 19.95& 19.66& 18.34& 17.59& 17.15& 16.91& 16.78 \\
18.5& 23.30& 22.23& 21.65& 20.94& 20.32& 20.03& 18.67& 17.89& 17.48& 17.26& 17.12 \\
19.0& 23.76& 22.66& 22.06& 21.35& 20.71& 20.42& 19.03& 18.26& 17.93& 17.74& 17.61 \\
\hline
&&&&${\rm log}_{10}$&$(n_{1/2}$&$/{\rm cm^{-3}})$ \\
\hline 
14.0&  3.29&  3.49&  3.67&  3.97&  4.30&  4.46&  5.17&  5.47&  5.53&  5.30&  4.70\\
14.5&  3.27&  3.48&  3.66&  3.96&  4.30&  4.45&  5.16&  5.47&  5.53&  5.30&  4.70\\
15.0&  3.22&  3.45&  3.64&  3.94&  4.29&  4.45&  5.16&  5.47&  5.53&  5.30&  4.70\\
15.5&  3.07&  3.34&  3.56&  3.89&  4.26&  4.42&  5.15&  5.46&  5.52&  5.30&  4.70\\
16.0&  2.72&  3.09&  3.35&  3.74&  4.16&  4.34&  5.13&  5.45&  5.51&  5.29&  4.68\\
16.5&  2.24&  2.65&  2.95&  3.42&  3.92&  4.14&  5.06&  5.41&  5.48&  5.26&  4.64\\
17.0&  1.74&  2.15&  2.47&  2.95&  3.49&  3.74&  4.86&  5.30&  5.39&  5.17&  4.53\\
17.5&  1.24&  1.65&  1.97&  2.45&  3.00&  3.25&  4.47&  5.02&  5.16&  4.94&  4.27\\
18.0&  0.742&  1.15&  1.47&  1.95&  2.50&  2.75&  3.98&  4.57&  4.73&  4.52&  3.84\\
18.5&  0.242& 0.652&  0.966&  1.45&  2.00&  2.25&  3.48&  4.07&  4.24&  4.03&  3.35\\
19.0& -0.258&  0.152&  0.466&  0.954&  1.50&  1.75&  2.98&  3.57&  3.74&  3.53&  2.85\\
\hline
&&&& $\alpha$&&&& \\
\hline 
14.0& 0.439&  0.409&  0.392&  0.370&  0.361&  0.357&  0.385&  0.437&  0.428&  0.354&  0.322\\
14.5&  0.436&  0.407&  0.391&  0.368&  0.359&  0.356&  0.385&  0.437&  0.427&  0.354&  0.322\\
15.0&  0.428&  0.401&  0.385&  0.364&  0.356&  0.352&  0.383&  0.436&  0.427&  0.352&  0.320\\
15.5&  0.416&  0.388&  0.373&  0.353&  0.347&  0.345&  0.380&  0.434&  0.425&  0.349&  0.316\\
16.0&  0.416&  0.378&  0.360&  0.338&  0.332&  0.330&  0.371&  0.429&  0.421&  0.341&  0.307\\
16.5&  0.450&  0.396&  0.367&  0.334&  0.322&  0.317&  0.355&  0.419&  0.414&  0.329&  0.292\\
17.0&  0.492&  0.435&  0.403&  0.362&  0.339&  0.329&  0.343&  0.406&  0.401&  0.317&  0.276\\
17.5&  0.529&  0.473&  0.441&  0.404&  0.381&  0.370&  0.362&  0.410&  0.392&  0.316&  0.272\\
18.0&  0.555&  0.503&  0.473&  0.440&  0.423&  0.414&  0.418&  0.446&  0.404&  0.335&  0.289\\
18.5&  0.582&  0.528&  0.499&  0.469&  0.457&  0.451&  0.470&  0.487&  0.432&  0.364&  0.310\\
19.0&  0.596&  0.546&  0.519&  0.492&  0.483&  0.479&  0.510&  0.516&  0.448&  0.372&  0.313\\
\hline
\end{tabular}
\end{center}
\caption{Rotational cooling parameters for CO}
\label{table:CO}
\end{table}

\begin{table}
\begin{center}
\renewcommand{\arraystretch}{1.0}
\begin{tabular}{ccccccc} \hline
${\rm log_{10}}(\tilde{N})$ &&&& temperature \\
\hline
 & 30K & 50K & 80K & 100K & 300K & 600K \\
\hline
&&$-{\rm log}_{10}$&$(L_{0}$ &$/{\rm ergs~cm^3~s^{-1}})$\\
\hline  
&25.31& 24.53& 24.02& 23.82& 23.16& 22.90\\
\hline
&& $-{\rm log}_{10}$&$({\cal L_{\rm LTE}}$&$/{\rm ergs~cm^3~s^{-1}})$ \\
\hline 
10.0& 16.20& 15.44& 14.90& 14.67& 13.81& 13.58\\
11.0& 16.20& 15.45& 14.90& 14.67& 13.81& 13.58\\
12.0& 16.22& 15.46& 14.91& 14.67& 13.81& 13.59\\
13.0& 16.36& 15.55& 14.96& 14.71& 13.83& 13.59\\
14.0& 17.02& 16.00& 15.25& 14.95& 13.94& 13.67\\
15.0& 17.74& 16.60& 15.77& 15.45& 14.49& 14.15\\
16.0& 18.58& 17.38& 16.50& 16.16& 15.10& 14.77\\
17.0& 19.42& 18.28& 17.40& 17.05& 15.84& 15.40\\
18.0& 20.38& 19.21& 18.36& 18.02& 16.83& 16.35\\
\hline
&&${\rm log}_{10}$&$(n_{1/2}$&$/{\rm cm^{-3}})$ \\
\hline
10.0&  9.05&  8.98&  8.90&  8.88&  8.77&  8.72\\
11.0&  9.05&  8.98&  8.90&  8.87&  8.77&  8.71\\
12.0&  9.03&  8.97&  8.89&  8.87&  8.76&  8.71\\
13.0&  8.88&  8.85&  8.81&  8.80&  8.74&  8.69\\
14.0&  8.20&  8.27&  8.36&  8.40&  8.50&  8.50\\
15.0&  7.25&  7.45&  7.71&  7.79&  7.99&  8.01\\
16.0&  6.26&  6.46&  6.75&  6.85&  7.22&  7.29\\
17.0&  5.26&  5.46&  5.75&  5.85&  6.23&  6.33\\
18.0&  4.26&  4.46&  4.75&  4.85&  5.23&  5.33\\
\hline
&& $\alpha$ \\
\hline
10.0&  0.353&  0.543&  0.536&  0.530&  0.466&  0.434\\
11.0&  0.354&  0.543&  0.536&  0.530&  0.466&  0.434\\
12.0&  0.364&  0.544&  0.536&  0.529&  0.466&  0.434\\
13.0&  0.399&  0.556&  0.534&  0.525&  0.463&  0.432\\
14.0&  0.655&  0.586&  0.532&  0.521&  0.447&  0.424\\
15.0&  0.641&  0.538&  0.538&  0.538&  0.448&  0.443\\
16.0&  0.649&  0.572&  0.552&  0.516&  0.381&  0.370\\
17.0&  0.566&  0.618&  0.580&  0.538&  0.381&  0.363\\
18.0&  0.701&  0.645&  0.598&  0.550&  0.383&  0.369\\
 \end{tabular}
\end{center}
\caption{Rotational cooling parameters for OH}
\label{table:OH}
\end{table}
\end{document}